\documentclass[useAMS,usenatbib]{mn2e}
\usepackage{times}
\usepackage{booktabs}
\usepackage{dcolumn}

\usepackage{amsfonts}
\usepackage{amssymb}
\usepackage{amsmath}
\usepackage{graphicx}
\usepackage{multirow}
\usepackage{subfigure}
\usepackage{times}
\usepackage{url}

\def\aj{AJ}
\def\araa{ARA\&A}
\def\apj{ApJ}
\def\apjl{ApJ}
\def\apjs{ApJS}

\def\apss{Ap\&SS}
\def\aap{A\&A}

\def\aaps{A\&AS}

\def\mnras{MNRAS}

\def\pasa{PASA}
\def\pasp{PASP}

\def\physscr{Phys.~Scr}

\def\rmxaa{RMxAA}

\def\arcmin{\hbox{$^\prime$}}
\def\arcsec{\hbox{$^{\prime\prime}$}}

\newcommand{\foiii}{[O~{\sc iii}]}

\newcommand{\foii}{[O~{\sc ii}]}

\newcommand{\fsii}{[S~{\sc ii}]}
\newcommand{\fsiii}{[S~{\sc iii}]}

\newcommand{\fnii}{[N~{\sc ii}]}

\newcommand{\fariii}{[Ar~{\sc iii}]}

\newcommand{\fneiii}{[Ne~{\sc iii}]}

\newcommand{\cii}{C~{\sc ii}}

\newcommand{\hi}{H\,{\sc i}}

\newcommand{\hei}{He~{\sc i}}

\def\p0{\phantom{0}}

\def\lessim{\raise-.5ex\hbox{$\buildrel<\over{\scriptstyle\mathtt{\sim}}$}}
\def\grtsim{\raise-.5ex\hbox{$\buildrel>\over{\scriptstyle\mathtt{\sim}}$}}

\title[The Wolf--Rayet planetary nebula Abell 48]{Observations and three-dimensional photoionization modelling of the Wolf--Rayet planetary nebula Abell~48\thanks{Based on observations made with the Australian National University (ANU) Telescope at the Siding Spring Observatory,  
and the Southern African Large Telescope (SALT) under
programs \mbox{2010-3-RSA\_OTH-002}.
}}
\author[A.~Danehkar et al.]{A.~Danehkar,$^{\,1}$\thanks{E-mail: ashkbiz.danehkar@students.mq.edu.au} 
H.~Todt,$^{\,2}$
B.~Ercolano$^{\,3,4}$ and
A.\,Y.~Kniazev$^{\,5,6,7}$
\\
$^{1}$Department of Physics and Astronomy, Macquarie University, Sydney, NSW 2109, Australia\\
$^{2}$Institut f\"{u}r Physik und Astronomie, Universit\"{a}t Potsdam, Karl-Liebknecht-Str.\,24/25, D-14476 Potsdam, Germany\\ 
$^{3}$Universit\"{a}ts-Sternwarte M\"{u}nchen, Ludwig-Maxmilians Universit\"{a}t M\"{u}nchen, Scheinerstr.\,1, D-81679 M\"{u}nchen, Germany\\
$^{4}$Exzellenzcluster Universe, Technische Universit\"{a}t M\"{u}nchen, Boltzmannstr.\,2, D-85748 Garching, Germany\\
$^{5}$South African Astronomical Observatory, PO Box 9, 7935 Observatory, Cape Town, South Africa \\
$^{6}$Southern African Large Telescope Foundation, PO Box 9, 7935 Observatory, Cape Town, South Africa \\
$^{7}$Sternberg Astronomical Institute, Lomonosov Moscow State University, Moscow 119992, Russia\\                             
}

\begin{document}

\date{Accepted 2014 January 28. Received 2014 January 28; in original form 2013 September 10}

\pagerange{\pageref{firstpage}--\pageref{lastpage}} \pubyear{2014}

\maketitle

\begin{abstract}Recent observations reveal that the central star of the planetary nebula Abell~48 exhibits spectral features similar to massive nitrogen-sequence Wolf--Rayet stars. This raises a pertinent question, whether it is still a planetary nebula or rather a ring nebula of a massive star. In this study, we have constructed a three-dimensional photoionization model of Abell~48, constrained by our new optical integral field spectroscopy. An analysis of the spatially resolved velocity distributions allowed us to constrain the geometry of Abell~48.  We used the collisionally excited lines to obtain the nebular physical conditions and ionic abundances of nitrogen, oxygen, neon, sulphur and argon, relative to hydrogen. We also determined helium temperatures and ionic abundances of helium and carbon from the optical recombination lines. We obtained a good fit to the observations for most of the emission-line fluxes in our photoionization model. The ionic abundances deduced from our model are in decent agreement with those derived by the empirical analysis.  However, we notice obvious discrepancies between helium temperatures derived from the model and the empirical analysis, as overestimated by our model. This could be due to the presence of a small fraction of cold metal-rich structures, which were not included in our model. It is found that the observed nebular line fluxes were best reproduced by using a hydrogen-deficient expanding model atmosphere as the ionizing source with an effective temperature of $T_{\rm eff}$~=~70\,kK and a stellar luminosity of $L_{\rm \star}$~=~5500\,L$_{\bigodot}$, which corresponds to a relatively low-mass progenitor star ($\sim3$~M$_{\bigodot}$) rather than a massive Pop I star. 
 
\end{abstract}

\label{firstpage}


\section{Introduction}
\label{a48:sec:introduction}

The highly reddened planetary nebula Abell~48 (PN\,G029.0$+$00.4) and its central star (CS) have been the subject of recent spectroscopic studies \citep{Wachter2010,Depew2011,Todt2013,Frew2013}. The CS of Abell\,48 has been classified as Wolf--Rayet [WN5] \citep{Todt2013}, where the square brackets distinguish it from the massive WN stars. 
Abell 48 was first identified as a planetary nebula (PN) by \citet{Abell1955}. However, its nature remains a source of controversy whether it is a massive ring nebula or a PN as previously identified. Recently,  \citet{Wachter2010} described it as a spectral type of WN6 with a surrounding ring nebula. But, \citet{Todt2013} concluded from spectral analysis of the CS and the surrounding nebula that Abell 48 is rather a PN with a low-mass CS than a massive (Pop I) WN star. Previously, \citet{Todt2010} also associated the CS of PB\,8 with [WN/C] class. Furthermore, IC\,4663 is another PN found to possess a [WN] star \citep{Miszalski2012}. 

A narrow-band H$\alpha$+[N\,\textsc{ii}] image of Abell~48 obtained by \citet{Jewitt1986} first showed its faint double-ring morphology. \citet{Zuckerman1986} identified it as a member of the elliptical morphological class. The H$\alpha$ image obtained from the SuperCOSMOS Sky H$\alpha$ Survey \citep{Parker2005} shows that the angular dimensions of the shell are about 46$\arcsec \times$ 38$\arcsec$, and are used throughout this paper. The first integral field spectroscopy of Abell~48 shows the same structure in the H$\alpha$ emission-line profile. But, a pair of bright point-symmetric regions is seen in [N\,\textsc{ii}] (see Fig.\,\ref{a48:fig3}), which could be because of the N$^{+}$ stratification layer produced by the photoionization process. A detailed study of the kinematic and ionization structure has not yet been carried out to date. This could be due to the absence of spatially resolved observations. 

\begin{table}
\caption{Journal of the IFU observations with the ANU 2.3-m Telescope.}
\label{a48:tab:obs:journal}
\centering
\begin{tabular}{lcccc}
\hline
\hline
PN & Date ({\sc ut})   & $\lambda$ range ({\AA}) & $R$ & Exp.(s) \\
\hline
Abell 48 &  2010/04/22  & 4415--5589 & 7000 & 1200\\
         &              & 5222--7070 & 7000 & 1200\\
         &  2012/08/23  & 3295--5906 & 3000 & 1200\\
         &              & 5462--9326 & 3000 & 1200\\
\hline 
\end{tabular}
\end{table}

The main aim of this study is to investigate whether the [WN] model atmosphere from \citet{Todt2013} of a low-mass star can reproduce the ionization structure of a PN with the features like Abell~48. We present integral field unit (IFU) observations and a three-dimensional photoionization model of the ionized gas in Abell~48. The paper is organized as follows. Section \ref{a48:sec:observations} presents our new observational data. In Section~\ref{a48:sec:kinematic} we describe the morpho-kinematic structure, followed by an empirical analysis in Section~\ref{a48:sec:empirical}. We describe our photoionization model and the derived results in Sections~\ref{a48:sec:photoionization} and \ref{a48:sec:modelresults}, respectively. Our final conclusion is stated in Section\,\ref{a48:sec:conclusions}.

\section{Observations and data reduction}
\label{a48:sec:observations}

\renewcommand{\baselinestretch}{0.9}
\begin{figure}
\begin{center}
\includegraphics[width=1.64in]{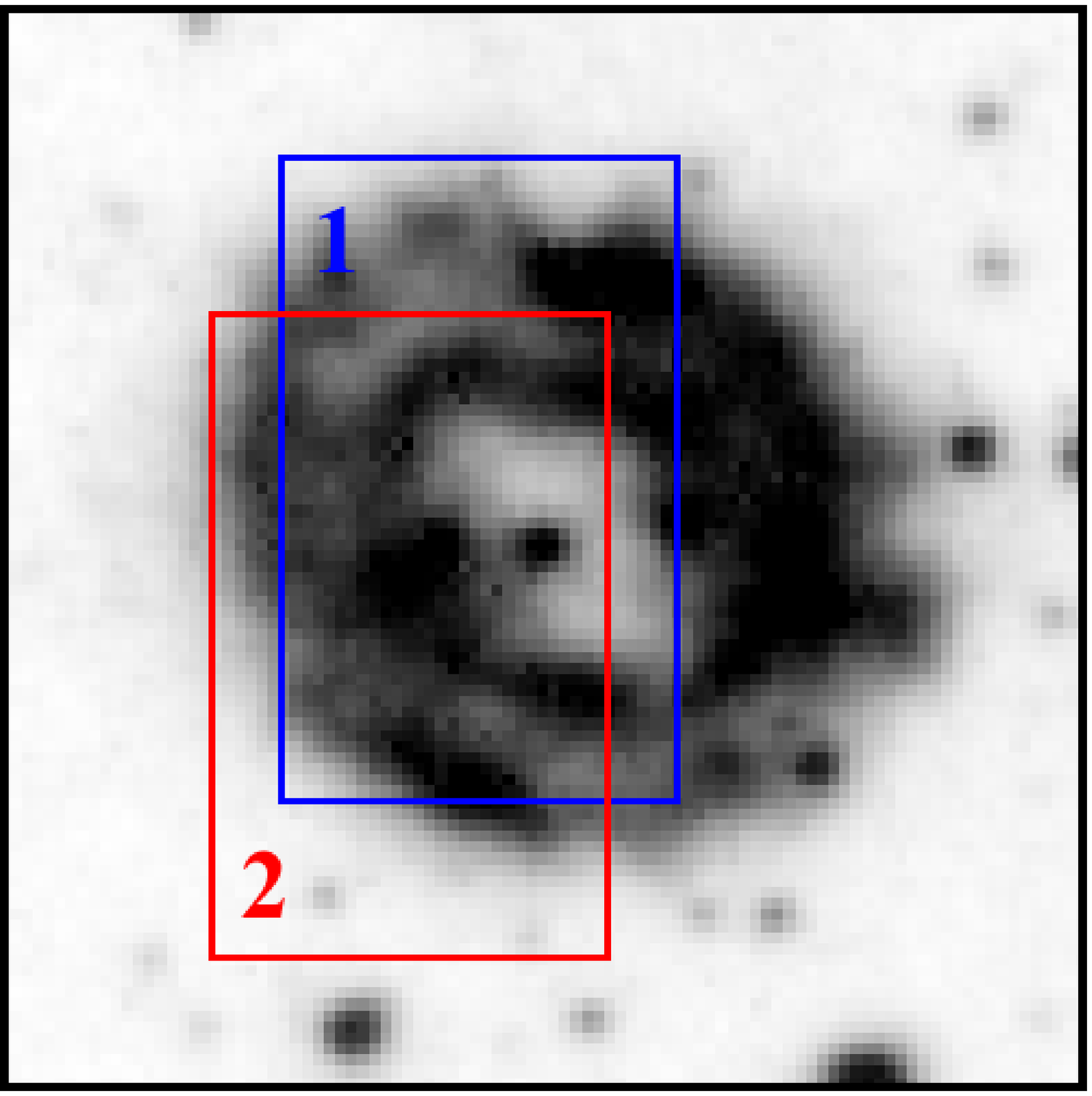}%
\includegraphics[width=1.7in]{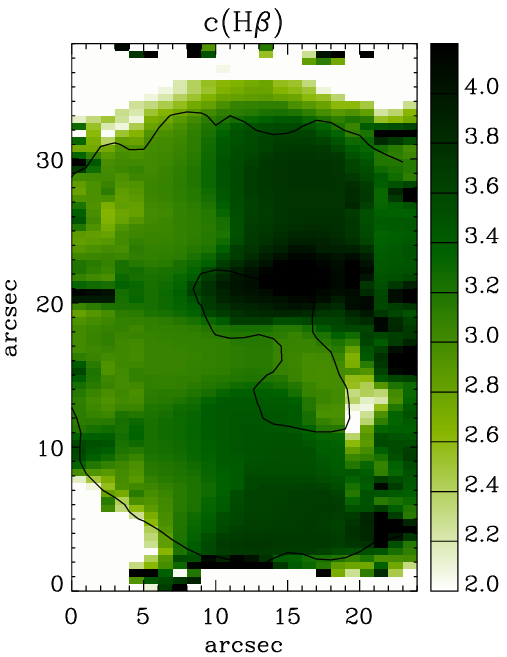}%
\hspace{32 mm}(a)\hspace{38 mm}(b)
\caption{From left to right: (a) narrow-band filter image of PN~Abell~48 in 
H$\alpha$ obtained from the SuperCOSMOS
Sky H$\alpha$ Survey \citep[SHS;][]{Parker2005}. The rectangles correspond the $25 \times 38$-arcsec${}^{2}$ IFU: 1 (blue) and 2 (red) taken in 2010 April and 2012 August, respectively. Image dimension is $60 \times 60$ arcsec${}^{2}$. 
(b) Extinction $c({\rm H}\beta)$ map of Abell\,48 calculated from the flux ratio H$\alpha$/H$\beta$ from fields. 
Black contour lines show the distribution of the narrow-band emission of H$\alpha$ in arbitrary unit obtained from the SHS. North is up and east is towards the left-hand side. 
\label{a48:fig1}%
}%
\end{center}
\end{figure}
\renewcommand{\baselinestretch}{1.5}

Integral field spectra listed in Table~\ref{a48:tab:obs:journal} were obtained in 2010 and 2012 with the 2.3-m ANU telescope using the Wide Field Spectrograph \citep[WiFeS;][]{Dopita2007,Dopita2010}. The observations were done with a spectral resolution of $R\sim 7000$ in the 441.5--707.0 nm range in 2010 and $R\sim 3000$ in the 329.5--932.6 nm range in 2012. The WiFeS has a field-of-view of $25\arcsec \times 38\arcsec$  and each spatial resolution element of  $1\farcs0\times0\farcs5$ (or $1\arcsec \times 1\arcsec$). The spectral resolution of $R\,(=\lambda/\Delta\lambda)\sim3000$ and $R\sim7000$ corresponds to a full width at half-maximum (FWHM) of $\sim100$ and 45 km\,s${}^{-1}$, respectively. We used the classical data accumulation mode, so a suitable sky window has been selected from the science data for the sky subtraction purpose. 

\begin{table}
\caption{\label{a48:tab:observations}Observed and dereddened relative line fluxes of the PN Abell~48, on a scale where H$\beta=100$.
Uncertain and very uncertain values are followed by `:' and `::', respectively. 
The symbol `*' denotes blended emission lines. 
}
\centering
\begin{tabular}{lccccc}
\hline\hline
\noalign{\smallskip}
$\lambda_{\rm lab}$({\AA}) & ID &  Mult & $F(\lambda)$ &  $I(\lambda)$ & Err(\%) \\
\noalign{\smallskip}
\hline
\noalign{\smallskip}
$   3726.03$ &    \foii &         F1 & $     20.72$: & $    128.96$:  & $ 25.7$\\
$   3728.82$ &    \foii &         F1 &           *  &           *   &      * \\
$   3868.75$ &  \fneiii &         F1 & $      7.52$ & $     38.96$  & $  9.4$\\
$   4340.47$ &    \hi\,5-2 &         H5 & $     21.97$ & $     54.28$:  & $  17.4$\\
$   4471.50$ &      \hei &        V14 & $      3.76$: & $      7.42$:  & $ 12.0$\\
$   4861.33$ &    \hi\,4-2 &         H4 & $    100.00$ & $    100.00$  & $  6.2$\\
$   4958.91$ &   \foiii &         F1 & $    117.78$ & $     99.28$  & $  5.3$\\
$   5006.84$ &   \foiii &         F1 & $    411.98$ & $    319.35$  & $  5.2$\\
$   5754.60$ &    \fnii &         F3 & $      1.73$:: & $      0.43$::  & $ 40.8$\\
$   5875.66$ &      \hei &        V11 & $     87.70$ & $     18.97$  & $  5.3$\\
$   6312.10$ &   \fsiii &         F3 & $      4.47$:: & $      0.60$::  & $ 46.9$\\
$   6461.95$ &      \cii &     V17.04 & $      3.36$: & $      0.38$:  & $ 26.2$\\
$   6548.10$ &    \fnii &         F1 & $    252.25$ & $     26.09$  & $  5.2$\\
$   6562.77$ &   \hi\,3-2  &         H3 & $   2806.94$ & $    286.00$  & $  5.1$\\
$   6583.50$ &    \fnii &         F1 & $    874.83$ & $     87.28$  & $  5.3$\\
$   6678.16$ &      \hei &        V46 & $     55.90$ & $      5.07$  & $  5.3$\\
$   6716.44$ &    \fsii &         F2 & $     85.16$ & $      7.44$  & $  5.1$\\
$   6730.82$ &    \fsii &         F2 & $     92.67$ & $      7.99$  & $  5.5$\\
$   7135.80$ &  \fariii &         F1 & $    183.86$ & $     10.88$  & $  5.2$\\
$   7236.42$ &      \cii &         V3 & $     29.96$: & $      1.63$:  & $ 20.7$\\
$   7281.35$ &      \hei &        V45 & $     11.08$:: & $      0.58$::  & $ 41.3$\\
$   7751.43$ &  \fariii &         F1 & $    111.83$:: & $      4.00$::  & $ 34.5$\\
$   9068.60$ &   \fsiii &         F1 & $   1236.22$ & $     19.08$  & $  5.3$\\
\noalign{\smallskip}
\hline
\noalign{\smallskip}
$c({\rm H}\beta)$ &      &            &              & \multicolumn{2}{l}{~~$3.10 \pm  0.04$}  \\
\multicolumn{2}{l}{H$\beta$/10$^{-13}$\,$\frac{\rm erg}{{\rm cm}^2{\rm s}}$}
&         \multicolumn{2}{r}{$1.076 \pm 0.067$}  & \multicolumn{2}{l}{~~$1354.6 \pm 154.2$} \\
\noalign{\smallskip}
\hline
\end{tabular}
\end{table}

The positions observed on the PN are shown in Fig. \ref{a48:fig1}(a). The centre of the IFU was placed in two different positions in 2010 and 2012. The exposure time of 20\,min yields a signal-to-noise ratio of $S/N \gtrsim 10$ for the [O~{\sc iii}] emission line. Multiple spectroscopic standard stars were observed for the flux calibration purposes, notably Feige\,110 and 
EG\,274. As usual, series of bias, flat-field frames, arc lamp exposures, and wire frames were acquired for data reduction, flat-fielding, wavelength calibration and spatial calibration.

Data reductions were carried out using the \textsc{iraf} pipeline {\sc wifes} (version 2.0; 2011 Nov 21).\footnote{IRAF is distributed by NOAO, which is operated by AURA, Inc., under contract to the National Science Foundation.} The reduction involves three main tasks: WFTABLE, WFCAL and WFREDUCE. The \textsc{iraf} task WFTABLE converts the raw data files with the single-extension Flexible Image Transport System (FITS) file format to the Multi-Extension FITS file format, edits FITS file key headers, and makes file lists for reduction purposes. The \textsc{iraf} task WFCAL extracts calibration solutions, namely the master bias, the master flat-field frame (from QI lamp exposures), the wavelength calibration (from Ne--Ar or Cu--Ar arc exposures and reference arc) and the spatial calibration (from wire frames). The \textsc{iraf} task WFREDUCE applies the calibration solutions to science data, subtracts sky spectra, corrects for differential atmospheric refraction, and applies the flux calibration using observations of spectrophotometric standard stars. 

\renewcommand{\baselinestretch}{0.9}
\begin{figure*}
\begin{center}
\includegraphics[width=1.70in]{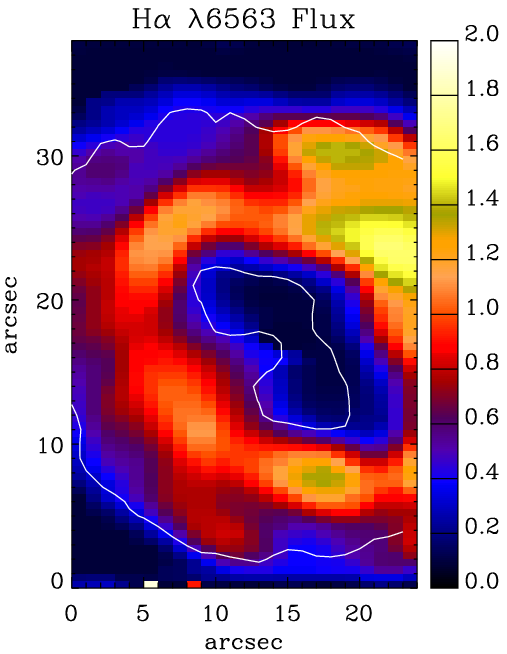}%
\includegraphics[width=1.70in]{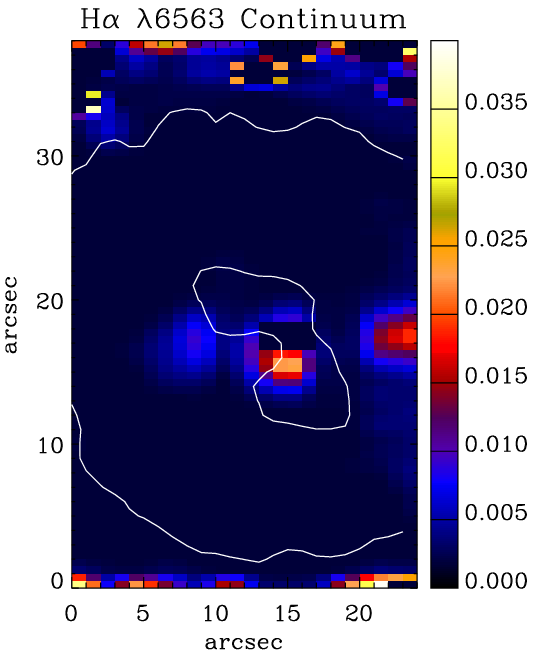}%
\includegraphics[width=1.70in]{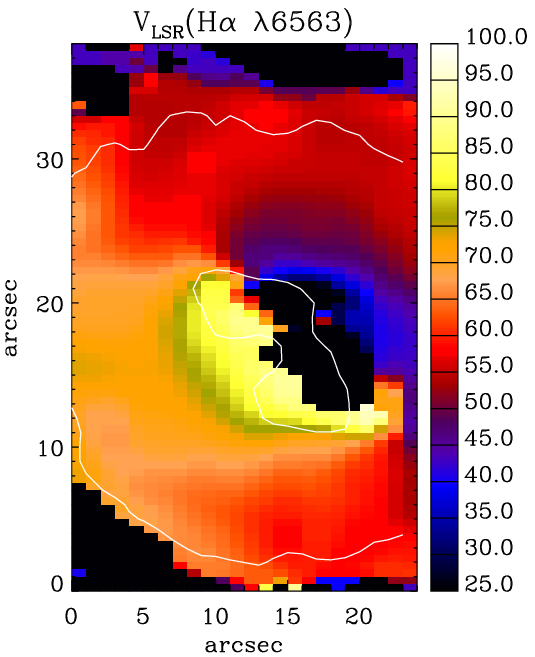}
\includegraphics[width=1.70in]{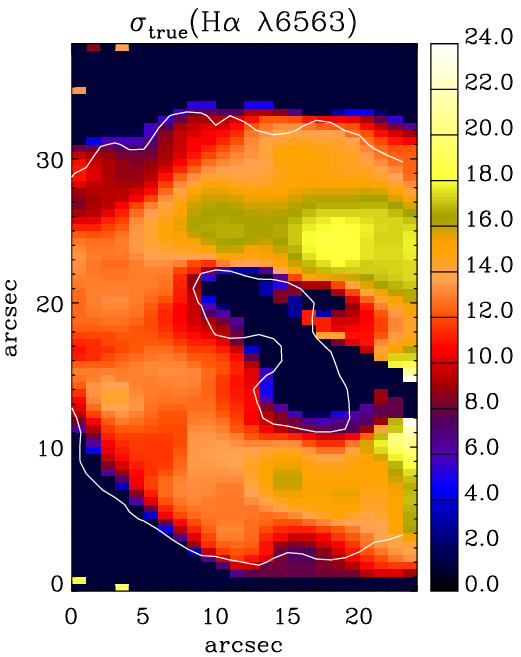}\\
\includegraphics[width=1.70in]{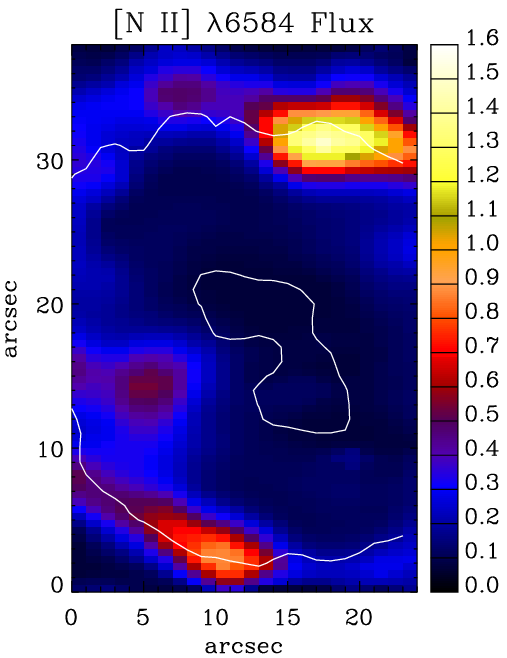}%
\includegraphics[width=1.70in]{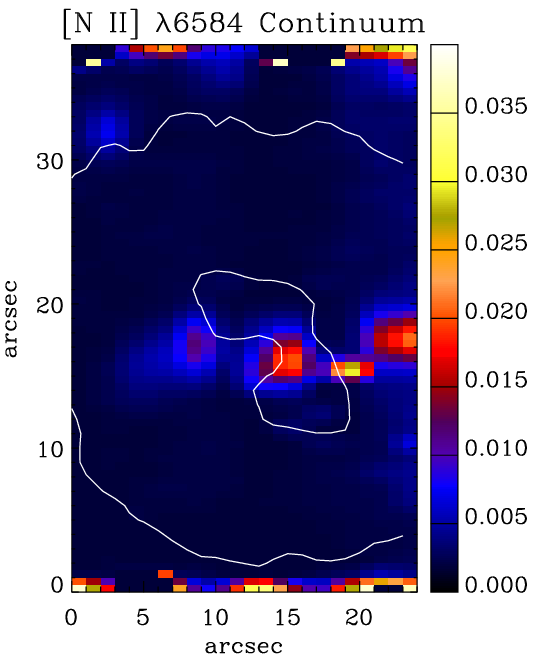}%
\includegraphics[width=1.70in]{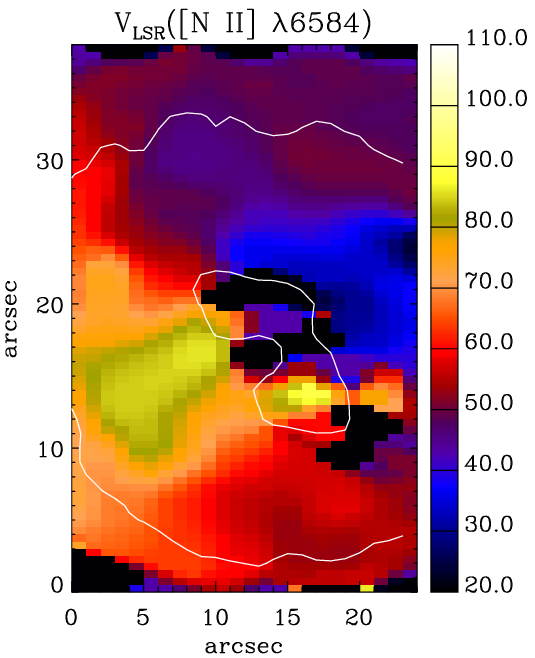}%
\includegraphics[width=1.70in]{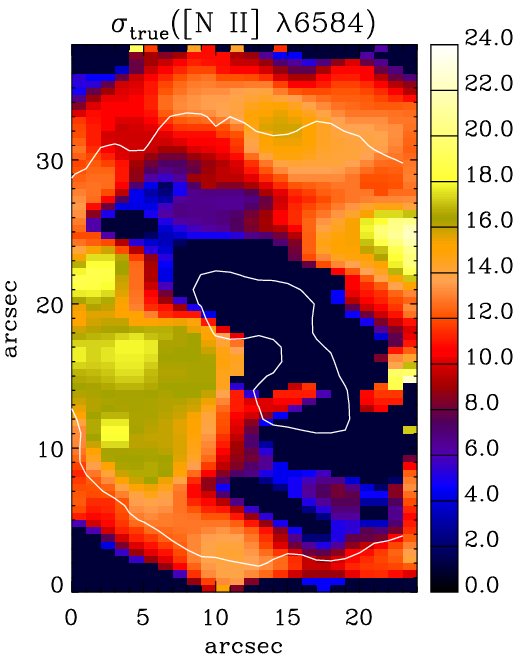}%
\caption{Maps of the PN Abell~48 in H$\alpha$ $\lambda$6563 {\AA} (top) and $[$N~{\sc ii}$]$ $\lambda$6584 {\AA} (bottom) from the IFU (${\rm PA}=0^{\circ}$) taken in 2010 April. From
left to right: spatial distribution maps of flux intensity, continuum, LSR velocity and velocity
dispersion. Flux unit is in $10^{-15}$~erg\,s${}^{-1}$\,cm${}^{-2}$\,spaxel${}^{-1}$, continuum in $10^{-15}$~erg\,s${}^{-1}$\,cm${}^{-2}$\,{\AA}${}^{-1}$\,spaxel${}^{-1}$, and velocities in km\,s${}^{-1}$. North is up and east is towards the left-hand side. 
The white contour lines show the distribution of the narrow-band emission of H$\alpha$ in arbitrary unit obtained from the SHS.
\label{a48:fig3}%
}%
\end{center}
\end{figure*}
\renewcommand{\baselinestretch}{1.5}

A complete list of observed emission lines and their flux intensities are given in Table~\ref{a48:tab:observations} on a scale where H$\beta$~=~100. All fluxes were corrected for reddening using 
$I(\lambda)_{\rm corr}=F(\lambda)_{\rm obs}10^{c({\rm H}\beta)[1+f(\lambda)]}.$
The logarithmic $c({\rm H}\beta)$ value of the interstellar extinction for the case B recombination \citep[$T_{\rm e}=10\,000$\,K and $N_{\rm e}=1000$\,cm$^{-3}$;][]{Storey1995} has been obtained from the H$\alpha$ and H$\beta$ Balmer fluxes. We used the Galactic extinction law $f(\lambda)$ of \citet{Howarth1983} for $R_V = A(V)/E(B-V)=3.1$, and normalized such that $f({\rm H}\beta)=0$. We obtained an extinction of $c({\rm H}\beta)=3.1$ for the total fluxes (see Table~\ref{a48:tab:observations}). Our derived nebular extinction is in excellent agreement with the value derived by \citet{Todt2013} from the stellar spectral energy (SED).  
The same method was applied to create $c({\rm H}\beta)$ maps using the flux ratio H$\alpha$/H$\beta$, as shown in Fig.~\ref{a48:fig1}(b).
Assuming that the foreground interstellar extinction is uniformly distributed over the nebula, an inhomogeneous extinction map may be related to some internal dust contributions. As seen, the extinction map of Abell\,48 depicts that the shell is brighter than other regions, and it may contain the asymptotic giant branch (AGB) dust remnants. 

\section{Kinematics}
\label{a48:sec:kinematic}

Fig.~\ref{a48:fig3} shows the spatial distribution maps of the flux intensity, continuum, radial velocity and velocity dispersion of H$\alpha$ $\lambda$6563 and $[$N~{\sc ii}$]$ $\lambda$6584 for Abell\,48. The white contour lines in the figures depict the distribution of the emission of H$\alpha$ obtained from the SHS \citep[][]{Parker2005}, which can aid us in distinguishing the nebular borders from the outside or the inside. The observed velocity $v_{\rm obs}$ was transferred to the local standard of rest (LSR) radial velocity $v_{\rm LSR}$ by correcting for the radial velocities induced by the motions of the Earth and Sun at the time of our observation. The transformation from the measured velocity dispersion $\sigma_{\rm obs}$ to the true line-of-sight velocity dispersion $\sigma_{\rm true}$ was done by 
$\sigma_{\rm true}=\sqrt{\sigma^2_{\rm obs}-\sigma^2_{\rm ins}-\sigma^2_{\rm th}}$, i.e. 
correcting for the instrumental width (typically $\sigma_{\rm ins}\approx42$\,km/s for $R\sim3000$ and $\sigma_{\rm ins}\approx18$\,km/s for $R\sim7000$) and the thermal broadening ($\sigma_{\rm th}^2=8.3\,T_{\rm e}[{\rm kK}]/Z$, where $Z$ is the atomic weight of the atom or ion). 

\renewcommand{\baselinestretch}{0.9}
\begin{figure}
\centering
{\footnotesize (a) Morpho-kinematic mesh model}\\
\includegraphics[width=3.20in]{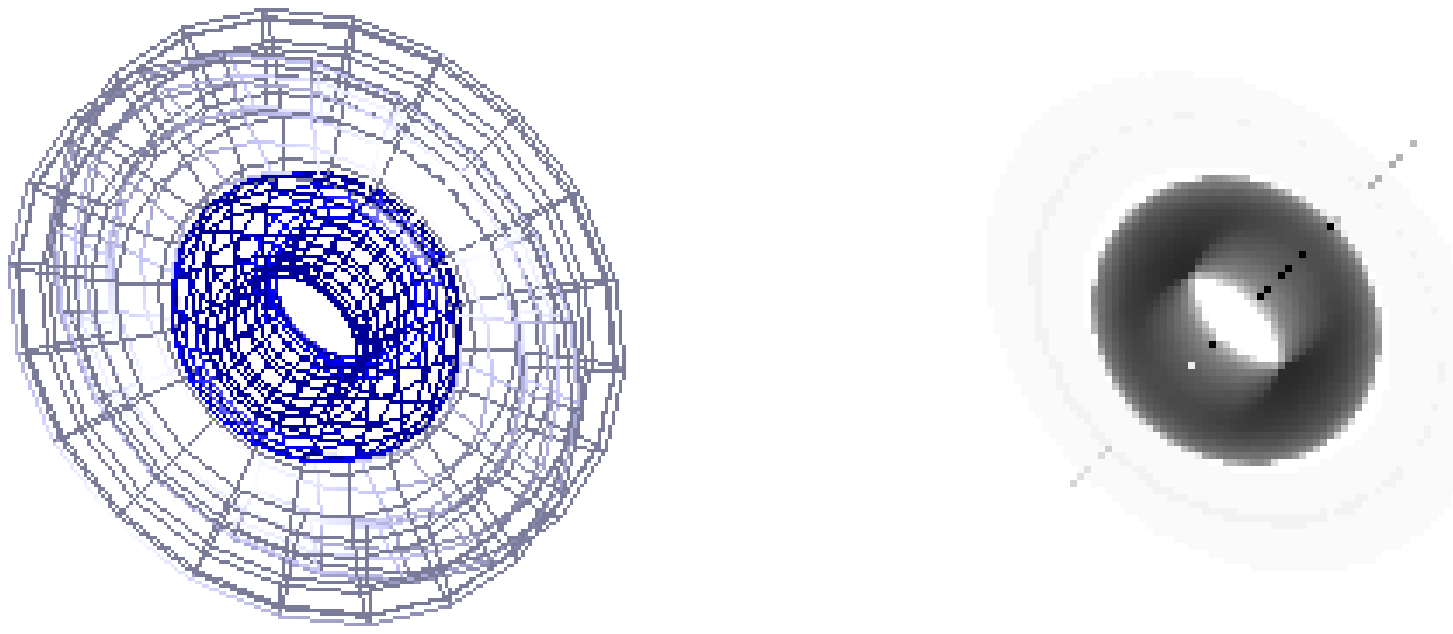}\\
{\footnotesize (b) Model results}\\
\includegraphics[width=1.65in]{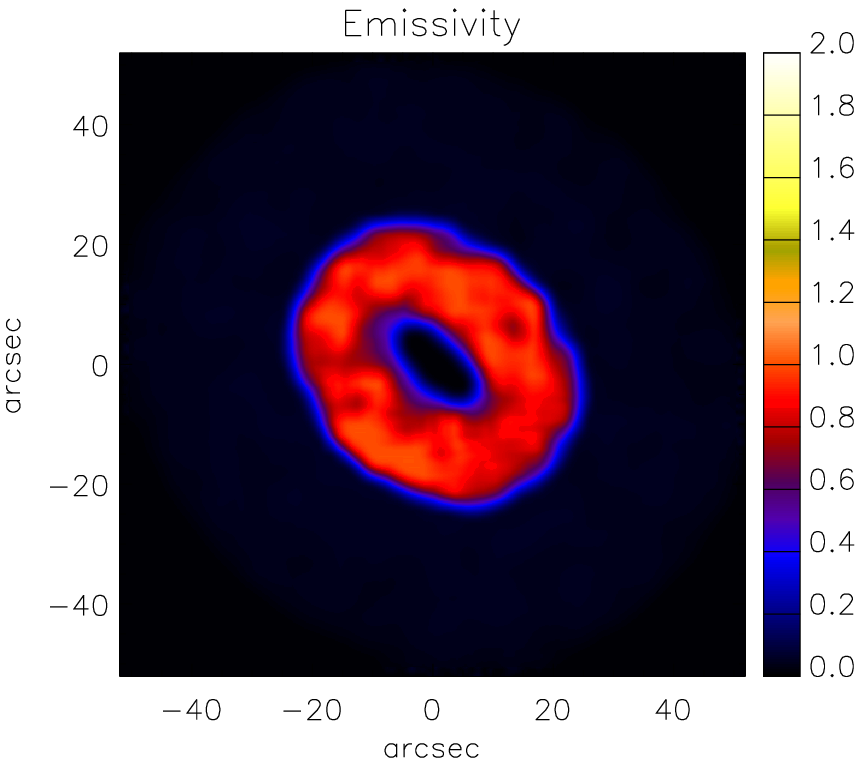}%
\includegraphics[width=1.65in]{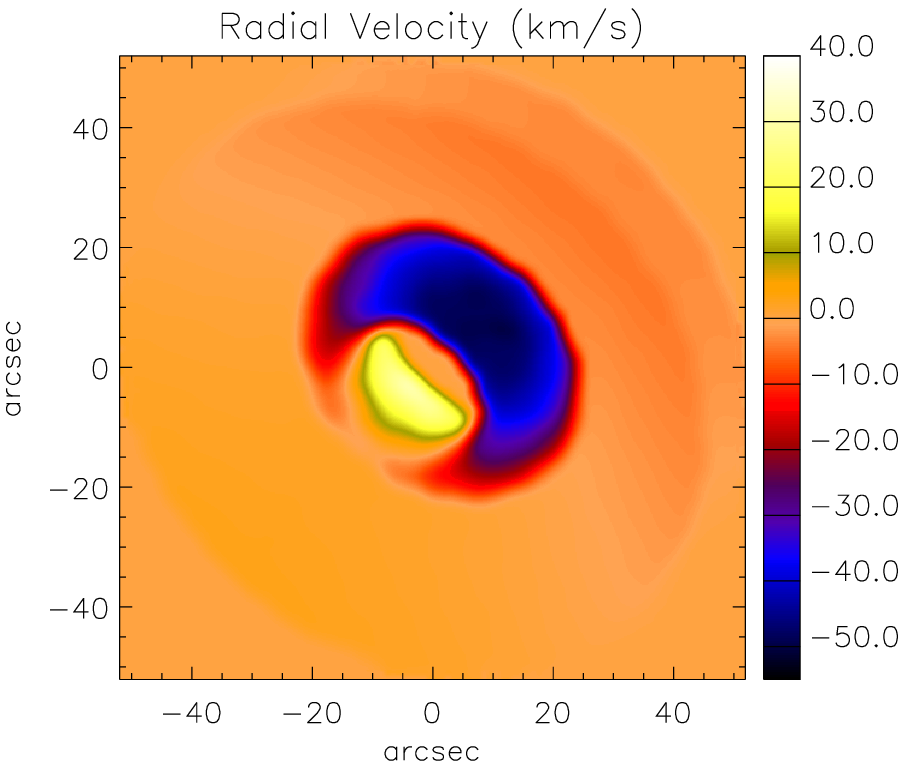}\\
\caption{(a) The \textsc{shape} mesh model before rendering at the best-fitting inclination and corresponding rendered model. (b) The normalized synthetic intensity map and the radial velocity map at the inclination 
of $-35{}^{\circ}$ and the position angle of $135{}^{\circ}$, derived from the model ($v_{\rm sys}=0$), which can be compared directly with Fig.~\ref{a48:fig3}. 
\label{a48:fig5}%
}%
\end{figure}
\renewcommand{\baselinestretch}{1.5}

We have used the three-dimensional morpho-kinematic modelling program \textsc{shape} (version 4.5) to study the kinematic structure. The program described in detail by \citet{Steffen2006} and \citet{Steffen2011}, uses interactively moulded geometrical polygon meshes to generate the 3D structure of objects. The modelling procedure consists of defining the geometry, emissivity distribution and velocity law as a function of position. The program produces several outputs that can be directly compared with long slit or IFU observations, namely the position--velocity (P--V) diagram, the 2-D line-of-sight velocity map on the sky and the projected 3-D emissivity on the plane of the sky. The 2-D line-of-sight velocity map on the sky can be used to interpret the IFU velocity maps. For best comparison with the IFU maps, the inclination ($i$), the position angle `\textit{PA}' in the plane of the sky, and the model parameters are modified in an iterative process until the qualitatively fitting 3D emission and velocity information are produced. We adopted a model, and then modified the geometry and inclination to conform to the observed H$\alpha$ and [N~{\sc ii}] intensity and radial velocity maps. For this paper, the three-dimensional structure has then been transferred to a regular cell grid, together with the physical emission properties, including the velocity that, in our case, has been defined as radially outwards from the nebular centre with a linear function of magnitude, commonly known as a Hubble-type flow \citep[see e.g.][]{Steffen2009}. 

The morpho-kinematic model of Abell\,48 is shown in Fig.~\ref{a48:fig5}(a), which consists of a modified torus, the nebular shell, surrounded by a modified hollow cylinder and the faint outer halo. The shell has an inner radius of $10\arcsec$ and an outer radius of $23\arcsec$ and a height of $23\arcsec$. We found an expansion velocity of $v_{\rm exp}=35\pm 5$\,km\,s${}^{-1}$ and a LSR  systemic velocity of $v_{\rm  sys}=65 \pm 5$\,km\,s${}^{-1}$. Our value of the LSR systemic velocity is in good agreement with the heliocentric systemic velocity of $v_{\rm hel}=50.4\pm4.2$\,km\,s${}^{-1}$ found by \citet{Todt2013}. Following \citet{Dopita1996}, we estimated the nebula's age around 1.5 of the dynamical age, so the star left the top of the AGB around $8880$ years ago.

Fig.~\ref{a48:fig5} shows the orientation of Abell\,48 on to the plane of the sky. The nebula has an inclination of $i=-35^{\circ}$ between the line of sight and the nebular symmetry axis. The symmetry axis has a position angle of ${\rm PA}=135^{\circ}$ projected on to the plane of the sky, measured from the north towards the east in the equatorial coordinate system  (ECS). The PA in the ECS can be transferred into the Galactic position angle (GPA) in the Galactic coordinate system (GCS), measured from the north Galactic pole (NGP; ${\rm GPA}=0^{\circ}$) towards the Galactic east (${\rm GPA}=90^{\circ}$). Note that ${\rm GPA}=90^{\circ}$ describes an alignment with the Galactic plane, while ${\rm GPA}=0^{\circ}$ is perpendicular to the Galactic plane. As seen in Table~\ref{a48:tab:kinematic:parameters}, Abell\,48 has a GPA of $197 \fdg 8$, meaning that the symmetry axis is approximately perpendicular to the Galactic plane. 

Based on the systemic velocity, Abell~48 must be located at less than 2\,kpc, since higher distances result in very high peculiar velocities \citep[$v_{\rm pec}>189$\,km\,s$^{-1}$; $v_{\rm pec}=170$\,km\,s$^{-1}$ found in few PNe in the Galactic halo by][]{Maciel1992}. However, it cannot be less than 1.5\,kpc due to the large interstellar extinction. 
Using the infrared dust maps\footnote{Website: \url{http://www.astro.princeton.edu/~schlegel/dust}} of \citet{Schlegel1998}, we found a mean reddening value of $E(B-V)=11.39 \pm  0.64$ for an aperture of $10 \arcmin$ in diameter in the Galactic latitudes and longitude of $(l,b)=(29.0,0.4)$, which is within a line-of-sight depth of $\lesssim20$\,kpc of the Galaxy. Therefore, Abell 48 with $E(B-V)\simeq2.14$ must have a distance of less than $3.3$ kpc. Considering the fact that the Galactic bulge absorbs photons overall 1.9 times more than the Galactic disc \citep{Driver2007}, the distance of Abell~48 should be around 2\,kpc, as it is located at the dusty Galactic disc.

\begin{table}
\caption{Kinematic results obtained for Abell\,48 based on the morpho-kinematic model matched to the observed 2-D radial velocity map.}
\label{a48:tab:kinematic:parameters}
\centering
\begin{tabular}{lc}
\hline
\hline
\noalign{\smallskip}
{Parameter\hspace{28 mm}}				& {Value} 							\\
\noalign{\smallskip}
\hline
\noalign{\smallskip}
$r_{\rm out}$ ({\scriptsize arcsec})  \dotfill   		&  $23 \pm 4 $ 		\\
$\delta r $ ({\scriptsize arcsec}) \dotfill  					&  $13 \pm 2$ 				\\
$h$ ({\scriptsize arcsec})\dotfill  							&  $23 \pm 4 $ 				\\
\noalign{\medskip}
$i$ \dotfill  							&  $-35^{\circ} \pm 2^{\circ}$		\\
PA \dotfill  						   	&  $135^{\circ} \pm 2^{\circ}$		\\
GPA \dotfill           &  $197^{\circ}48\arcmin \pm 2^{\circ} $	\\
\noalign{\medskip}
$v_{\rm sys}$({\scriptsize km/s}) \dotfill					&  $65 \pm 5$ 	\\
$v_{\rm exp}$({\scriptsize km/s}) \dotfill					&  $35\pm 5$ 	\\
\noalign{\medskip}
\hline
\end{tabular}
\end{table}

\section{Nebular empirical analysis}
\label{a48:sec:empirical}

\subsection{Plasma diagnostics}
\label{a48:sec:tempdens}

The derived electron temperatures ($T_{\rm e}$) and densities ($N_{\rm e}$) are listed in Table~\ref{ab48:tab:tenediagnostics}, together with the ionization potential required to create the emitting ions. 
We obtained $T_{\rm e}$ and $N_{\rm e}$ from temperature-sensitive and density-sensitive emission lines by solving the equilibrium equations of level populations for a multilevel atomic model using \textsc{equib} code \citep{Howarth1981}. The atomic data sets used for our plasma diagnostics from collisionally excited lines (CELs), as well as for abundances derived from CELs, are given in Table~\ref{a48:tab:atomicdata:cel}. The diagnostics procedure to determine temperatures and densities from CELs is as follows: we assume a representative initial electron temperature of 10\,000\,K in order to derive $N_{\rm e}$ from $[$S~{\sc ii}$]$ line ratio; then $T_{\rm e}$ is derived from $[$N~{\sc ii}$]$ line ratio in conjunction with the mean density derived from the previous step. The calculations are iterated to give self-consistent results for $N_{\rm e}$ and $T_{\rm e}$. The correct choice of electron density and temperature is important for the abundance determination.

We see that the PN Abell~48 has a mean temperature of $T_{\rm e}([$N~{\sc ii}$])=6980 \pm 930 $~K, and a mean electron density of $N_{\rm e}([$S~{\sc ii}$])=750 \pm 200$~cm${}^{-3}$, which are in reasonable agreement with $T_{\rm e}([$N~{\sc ii}$])=7\,200 \pm 750$~K and $N_{\rm e}([$S~{\sc ii}$])=1000 \pm 130$~cm${}^{-3}$ found by \citet{Todt2013}. The uncertainty on $T_{\rm e}([$N~{\sc ii}$])$ is order of $40$ percent or more, due to the weak flux intensity of [N~{\sc ii}] $\lambda$5755, the recombination contribution, and high interstellar extinction. Therefore, we adopted the mean electron temperature from our photoionization model for our CEL abundance analysis. 

\begin{table}
\footnotesize
\caption{References for atomic data.}
\label{a48:tab:atomicdata:cel}
\centering
\begin{tabular}{lll}
\hline\hline
\noalign{\smallskip}
Ion 	& Transition probabilities & Collision strengths \\
\noalign{\smallskip}
\hline 
\noalign{\smallskip} 
N${}^{+}$   & \citet{Bell1995}    & \citet{Stafford1994}  \\ 
\noalign{\medskip}
O${}^{+}$   & \citet{Zeippen1987} & \citet{Pradhan2006}  \\ 
O${}^{2+}$  & \citet{Storey2000}  &  \citet{Lennon1994} \\ 
\noalign{\medskip}
Ne${}^{2+}$ & \citet{Landi2005}   &   \citet{McLaughlin2000} \\ 
\noalign{\medskip}
S${}^{+}$   & \citet{Mendoza1982} & \citet{Ramsbottom1996}  \\ 
S${}^{2+}$  & \citet{Mendoza1982} & \citet{Tayal1999} \\ 
            & \citet{Huang1985}   &   \\  
\noalign{\medskip}
Ar${}^{2+}$ & \citet{Biemont1986} & \citet{Galavis1995} \\ 
\noalign{\smallskip}   
\hline
\noalign{\smallskip}
Ion 	& Recombination coefficient & Case \\
\noalign{\smallskip}
\hline 
\noalign{\smallskip} 
H${}^{+}$   & \citet{Storey1995} &  B \\ 
\noalign{\medskip}
He${}^{+}$   & \citet{Porter2013} &  B \\ 
\noalign{\medskip}
C${}^{2+}$  & \citet{Davey2000} &  B \\  
\noalign{\smallskip}   
\hline
\end{tabular}
\end{table}

Table~\ref{ab48:tab:tenediagnostics} also lists the derived He\,{\sc i} temperatures, which are lower than the CEL temperatures, known as the ORL-CEL temperature discrepancy problem in PNe \citep[see e.g.][]{Liu2000,Liu2004a}. 
To determine the electron temperature from the He\,{\sc i} $\lambda\lambda$5876, 6678 and 7281 lines, we used the emissivities of He~I lines by \citet{Smits1996}, which also include the temperature range of $T_{\rm e} < 5000$\,K. We derived electron temperatures of $T_{\rm e}({\rm He~I})=5110$\,K and $T_{\rm e}({\rm He~I})=4360$\,K from  the flux ratio He\,{\sc i} $\lambda\lambda$7281/5876 and $\lambda\lambda$7281/6678, respectively. Similarly, we got $T_{\rm e}({\rm He~I})=6960$\,K for  He\,{\sc i} $\lambda\lambda$7281/5876 and $T_{\rm e}({\rm He~I})=7510$\,K for $\lambda\lambda$7281/6678 from the measured nebular spectrum by \citet{Todt2013}. 

\begin{table}
\caption{Diagnostics for the electron temperature, $T_{\rm e}$ and the electron density, $N_{\rm e}$. References: D13 -- this work; T13 -- \citet{Todt2013}. }
\label{ab48:tab:tenediagnostics}
\centering
\begin{tabular}{lcccc}
\hline
\hline
\noalign{\smallskip}
{ Ion}   &  { Diagnostic}  & { I.P.(eV)}  &$T_{\rm e}({\rm K})$ & Ref. \\ 
\noalign{\smallskip}
\hline
\noalign{\smallskip}
$[$N~{\sc ii}$]$  & { $\frac{\lambda6548+\lambda6584}{\lambda5755}$}  &  14.53 & $6980 \pm 930$ & D13 \\
                  &                                                   &        & $7200 \pm 750$ & T13 \\
\noalign{\smallskip}
$[$O~{\sc iii}$]$  & { $\frac{\lambda4959+\lambda5007}{\lambda4363}$} &  35.12 & $11870 \pm 1640$ & T13 \\
\noalign{\smallskip}
He\,{\sc i}  & { $\frac{\lambda7281}{\lambda5876}$} &  24.59 & $5110 \pm  2320$ & D13\\
            &                                      &        & $6960 \pm 450$ & T13\\
\noalign{\smallskip}
He\,{\sc i}  & { $\frac{\lambda7281}{\lambda6678}$} &  24.59 & $4360 \pm  1820$ & D13 \\
            &                                      &        & $7510 \pm 4800$ & T13\\
\noalign{\smallskip}
\hline
\noalign{\smallskip}
 & & & $N_{\rm e}({\rm cm}^{-3})$ \\
\noalign{\smallskip}
\hline
\noalign{\smallskip}
$[$S~{\sc ii}$]$  & { $\frac{\lambda6717}{\lambda6731}$} & 10.36 & $750 \pm 200$ & D13 \\
\noalign{\smallskip}
                  &                                      &       & $1000 \pm 130$ & T13 \\
\noalign{\smallskip}
\hline
\end{tabular}
\end{table}

\subsection{Ionic and total abundances from ORLs}

Using the effective recombination coefficients (given in Table~\ref{a48:tab:atomicdata:cel}), we determine ionic abundances, X${}^{i+}$/H${}^{+}$, from the measured intensities of optical recombination lines (ORLs) as follows:
\begin{equation}
\frac{N({\rm X}^{i+})}{N({\rm H}^{+})}=\frac{I({\lambda})}{I({{\rm H}\beta})} 
\frac{\lambda({\rm {\AA}})}{4861} \frac{\alpha_{\rm eff}({\rm H}\beta)}{\alpha_{\rm eff}(\lambda)},
\label{a48:eq_orl1}%
\end{equation}
where $I({\lambda})$ is the intrinsic line flux of the emission line $\lambda$ emitted by ion ${\rm X}^{i+}$,  
$I({{\rm H}\beta})$ is the intrinsic line flux of H$\beta$, $\alpha_{\rm eff}({\rm H}\beta)$ the effective recombination coefficient of H$\beta$, and $\alpha_{\rm eff}(\lambda)$ the effective recombination coefficient for the emission line $\lambda$.

Abundances of helium and carbon from ORLs are given in Table~\ref{a48:tab:abundances:orls}. We derived the ionic and total helium abundances from He\,{\sc i} $\lambda$4471, $\lambda$5876 and $\lambda$6678 lines. We assumed the Case B recombination for the He\,{\sc i} lines \citep{Porter2012,Porter2013}. We adopted an electron temperature of $T_{\rm e}=5\,000$~K from He\,{\sc i} lines, and an electron density of $N_{\rm e}=1000$~cm${}^{-3}$. We averaged the He${}^{+}$/H${}^{+}$ ionic abundances from the He\,{\sc i} $\lambda$4471, $\lambda$5876 and $\lambda$6678 lines  with weights of 1:3:1, roughly the intrinsic intensity ratios of these three lines. The total He/H abundance ratio is obtained by simply taking the sum of He${}^{+}$/H${}^{+}$ and He${}^{2+}$/H${}^{+}$. However, He${}^{2+}$/H${}^{+}$ is equal to zero, since He\,{\sc ii} $\lambda$4686 is not present. The C$^{2+}$ ionic abundance is obtained from C~{\sc ii} $\lambda$6462 and $\lambda$7236 lines. 

\begin{table}
\caption{Empirical ionic abundances derived from ORLs.}
\label{a48:tab:abundances:orls}
\centering
\begin{tabular}{llcc}
\hline
\hline
\noalign{\smallskip}
Ion& { $\lambda$({\AA})} & {Mult} & {Value\,$^{\mathrm{a}}$ }  \\
\noalign{\smallskip}
\hline
\noalign{\smallskip}
He$^+$   &4471.50 & V14 & 0.141 \\                                      
         &5876.66 & V11 & 0.121 \\
         &6678.16 & V46 & 0.115 \\
         &Mean    &     & 0.124 \\
\noalign{\smallskip}
He$^{2+}$&4685.68 & 3.4 & 0.0 \\
\noalign{\smallskip}
He/H     &        &     & 0.124 \\
\noalign{\vskip8pt}                   
C$^{2+}$ &6461.95 & V17.40 & 3.068($-3$) \\  
		 &7236.42 &  V3    & 1.254($-3$)\\   
         &Mean    &        & 2.161($-3$) \\       
\noalign{\smallskip}
\hline 
\end{tabular}
\begin{list}{}{}
\item[$^{\mathrm{a}}$] Assuming $T_{\rm e}=5\,000$\,K and $N_{\rm e}=1000$\,${\rm cm}^{-3}$.
\end{list}
\end{table}

\subsection{Ionic and total abundances from CELs}

We determined abundances for ionic species of N, O, Ne, S and Ar from CELs. To deduce ionic abundances, we solve the statistical equilibrium equations for each ion using \textsc{equib} code, giving level population and line sensitivities for specified $N_{\rm e}=1000$~cm${}^{-3}$ and $T_{\rm e}=10\,000$~K adopted according to our photoionization modelling. Once 
the equations for the population numbers are solved, the ionic abundances, X${}^{i+}$/H${}^{+}$, can be derived from the observed line intensities of CELs as follows:
\begin{equation}
\frac{N({\rm X}^{i+})}{N({\rm H}^{+})}=\frac{I(\lambda_{ij})}{I({{\rm H}\beta})} 
\frac{\lambda_{ij}({\rm {\AA}})}{4861} \frac{\alpha_{\rm eff}({{\rm H}\beta})}{A_{ij}} 
\frac{N_{\rm e}}{n_i},
\label{a48:eq_cel1}
\end{equation}
where $I(\lambda_{ij})$ is the dereddened flux of the emission line $\lambda_{ij}$ emitted by ion ${\rm X}^{i+}$ following the transition from the upper level $i$ to the lower level $j$,  
$I({{\rm H}\beta})$ the dereddened flux of H$\beta$, $\alpha_{\rm eff}({{\rm H}\beta})$ the effective recombination coefficient of H$\beta$, $A_{ij}$ the Einstein spontaneous transition probability of the transition, $n_i$ the fractional population of the upper level $i$, and $N_{\rm e}$ is the electron density.

Total elemental  and ionic abundances of nitrogen, oxygen, neon, sulphur and argon from CELs are presented in Table~\ref{a48:tab:abundances:cels}. Total elemental abundances are derived from ionic abundances using the ionization correction factors ($icf$) formulas given by \citet{Kingsburgh1994}. The total O/H abundance ratio is obtained by simply taking the sum of the O$^{+}$/H$^{+}$ derived from [O~{\sc ii}] $\lambda\lambda$3726,3729 doublet, and the O$^{2+}$/H$^{+}$ derived from [O~{\sc iii}] $\lambda\lambda$4959,5007 doublet, since He\,{\sc ii} $\lambda$4686 is not present, so O${}^{3+}$/H${}^{+}$ is negligible. 
The total N/H abundance ratio was calculated from the N$^{+}$/H$^{+}$ ratio derived from the [N~{\sc ii}] $\lambda\lambda$6548,6584 doublet, correcting for the unseen N$^{2+}$/H$^{+}$ using,
\begin{equation}
\footnotesize
\frac{{\rm N}}{{\rm H}}=\left(\frac{{\rm N}^{+}}{{\rm H}^{+}}\right) \left(\frac{{\rm O}}{{\rm O}^{+}}\right).
\label{a48:icf:n_cel}
\end{equation}
The Ne$^{2+}$/H$^{+}$ is derived from [Ne~{\sc iii}] $\lambda$3869 line. Similarly, the unseen Ne$^{+}$/H$^{+}$ is corrected for, using
\begin{equation}
\frac{{\rm Ne}}{{\rm H}}=\left(\frac{{\rm Ne}^{2+}}{{\rm H}^{+}} \right)
\left(\frac{{\rm O}}{{\rm O}^{2+}}\right) .
\end{equation}
For sulphur, we have S$^{+}$/H$^{+}$ from the [S~{\sc ii}] $\lambda\lambda$6716,6731 doublet and 
S$^{2+}$/H$^{+}$ from the [S~{\sc iii}] $\lambda$9069 line. The total sulphur abundance is corrected for the unseen stages of ionization using
\begin{equation}
\footnotesize
\frac{{\rm S}}{{\rm H}}=\left(\frac{{\rm S}^{+}}{{\rm H}^{+}} + \frac{{\rm S}^{2+}}{{\rm H}^{+}} \right)
\left[1-\left(1-\frac{{\rm O}^{+}}{{\rm O}}\right)^{3}\right]^{-1/3}.
\label{a48:icf:s_cel}
\end{equation}
The [Ar~{\sc iii}] 7136 line is only detected, so we have only Ar$^{2+}$/H$^{+}$. The total argon abundance is obtained by assuming Ar$^{+}$/Ar~=~N$^{+}$/N:
\begin{equation}
\footnotesize
\frac{{\rm Ar}}{{\rm H}}=\left(\frac{{\rm Ar}^{2+}}{{\rm H}^{+}}  \right)
\left(1-\frac{{\rm N}^{+}}{{\rm N}}\right)^{-1}.
\label{a48:icf:cl_cel}
\end{equation}
As it does not include the unseen Ar$^{3+}$, so the derived elemental argon may be underestimated. 

\renewcommand{\baselinestretch}{0.9}
\begin{figure*}
\centering
\includegraphics[width=1.7in]{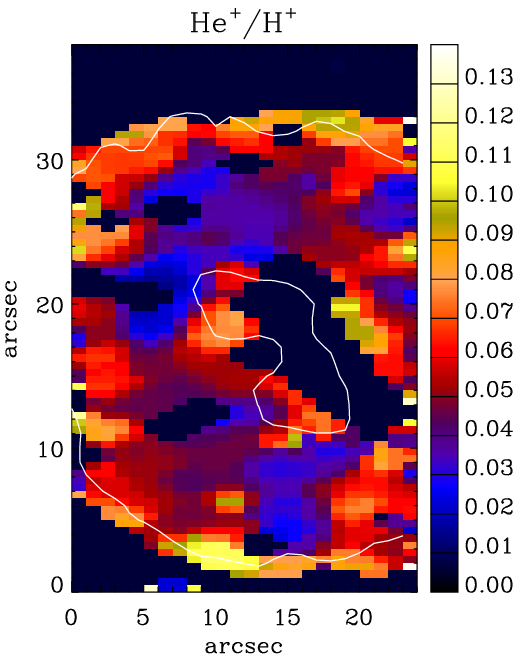}
\includegraphics[width=1.7in]{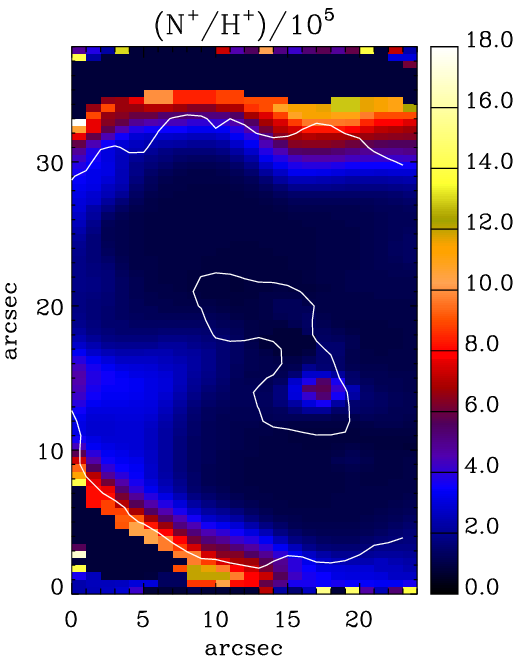}%
\includegraphics[width=1.7in]{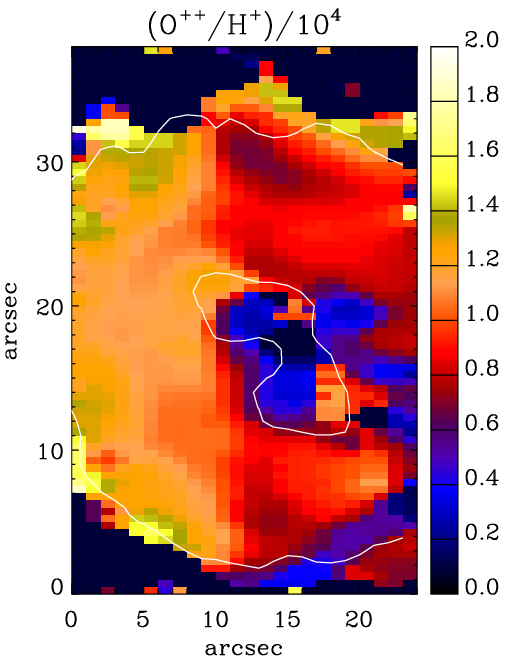}%
\includegraphics[width=1.7in]{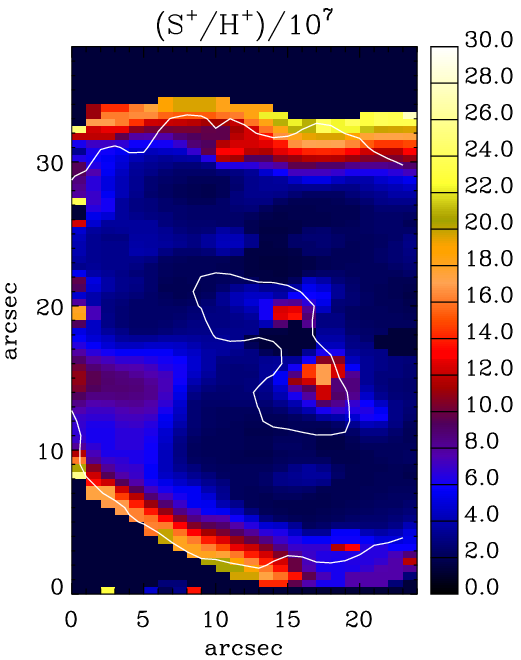}
\caption{Ionic abundance maps of Abell~48. From left to right: spatial distribution maps of singly ionized Helium abundance ratio He${}^{+}$/H${}^{+}$ from He\,{\sc i} ORLs (4472, 5877, 6678); ionic nitrogen abundance ratio N${}^{+}$/H${}^{+}$ ($\times 10^{-5}$) from $[$N~{\sc ii}$]$ CELs (5755, 6548, 6584); ionic oxygen abundance ratio O${}^{2+}$/H${}^{+}$ ($\times 10^{-4}$) from $[$O~{\sc iii}$]$ CELs (4959, 5007); and ionic sulphur abundance ratio S${}^{+}$/H${}^{+}$ ($\times 10^{-7}$) from $[$S~{\sc ii}$]$ CELs (6716, 6731). North is up and east is towards the left-hand side. The white contour lines show the distribution of the narrow-band emission of H$\alpha$ in arbitrary unit obtained from the SHS. 
\label{a48:ifu:ion:map}%
}%
\end{figure*}
\renewcommand{\baselinestretch}{1.5}

Fig.\,\ref{a48:ifu:ion:map} shows the spatial distribution of ionic abundance ratio He${}^{+}$/H${}^{+}$, N${}^{+}$/H${}^{+}$, O${}^{2+}$/H${}^{+}$ and S${}^{+}$/H${}^{+}$ derived for given $T_{\rm e}=10000$\,K and $N_{\rm e}=1000$\,cm$^{-3}$. We notice that both O${}^{2+}$/H${}^{+}$ and He${}^{+}$/H${}^{+}$ are very high over the shell, whereas  N${}^{+}$/H${}^{+}$ and S${}^{+}$/H${}^{+}$ are seen at the edges of the shell. It shows obvious results of the ionization sequence from the
highly inner ionized zones to the outer low ionized regions.

\renewcommand{\baselinestretch}{1.2}
\begin{table}
\caption{Empirical ionic abundances derived from CELs.}
\label{a48:tab:abundances:cels}
\centering
\begin{tabular}{llcc}
\hline
\hline
\noalign{\smallskip}
Ion& { $\lambda$({\AA})} & {Mult} & {Value\,$^{\mathrm{a}}$ }  \\
\noalign{\smallskip}
\hline    
\noalign{\smallskip}
N${}^{+}$ & 6548.10  & F1 & 1.356($-5$) \\ 
          & 6583.50  & F1 & 1.486($-5$) \\ 
          & Mean     &    & 1.421($-5$) \\ 
\noalign{\smallskip}
          & $icf$(N) &    & 3.026       \\
N/H       &          &    & 4.299($-5$) \\
\noalign{\vskip8pt} 
O${}^{+}$ & 3727.43  & F1 & 5.251($-5$) \\
\noalign{\smallskip} 
O${}^{2+}$ & 4958.91 & F1 & 1.024($-4$) \\ 
          & 5006.84  & F1 & 1.104($-4$) \\ 
          & Average  &    & 1.064($-4$) \\ 
\noalign{\smallskip}
          & $icf$(O) &    & 1.0 \\
O/H       &          &    & 1.589($-4$) \\
\noalign{\vskip8pt}  
Ne${}^{2+}$& 3868.75 & F1 & 4.256($-5$)  \\ 
\noalign{\smallskip}
           & $icf$(Ne)&    & 1.494 \\
Ne/H       &          &    & 6.358($-5$) \\  
\noalign{\vskip8pt}   
S${}^{+}$ & 6716.44 & F2 & 4.058($-7$) \\ 
          & 6730.82 & F2 & 3.896($-7$) \\ 
          & Average &    & 3.977($-7$) \\ 
\noalign{\smallskip}
S${}^{2+}$ & 9068.60 & F1 & 5.579($-6$) \\ 
\noalign{\smallskip}
          & $icf$(S) &    &  1.126 \\
S/H      &           &    & 6.732($-6$)  \\
\noalign{\vskip8pt}   
Ar${}^{2+}$& 7135.80 & F1 & 9.874($-7$) \\ 
\noalign{\smallskip}
          & $icf$(Ar)&    &  1.494 \\
Ar/H      &          &    &  1.475($-6$)  \\
\noalign{\smallskip}
\hline 
\end{tabular}
\begin{list}{}{}
\item[$^{\mathrm{a}}$] Assuming $T_{\rm e}=10\,000$\,K and $N_{\rm e}=1000$\,${\rm cm}^{-3}$.
\end{list}
\end{table}
\renewcommand{\baselinestretch}{1.5}

\section{Photoionization modelling}
\label{a48:sec:photoionization}

The 3-D photoionization code \textsc{mocassin} \citep[version 2.02.67;][]{Ercolano2003a,Ercolano2005,Ercolano2008} was used to study the best-fitting
model for Abell\,48. 
The code has been used to model a number of PNe, for example NGC~3918 \citep{Ercolano2003b}, NGC~7009 \citep{Gonccalves2006}, 
NGC~6302 \citep{Wright2011}, and SuWt~2 \citep{Danehkar2013}. The modelling procedure consists of defining the density distribution and elemental abundances of the nebula, as well as assigning the ionizing spectrum of the CS. This code uses a Monte Carlo method to solve self-consistently the 3-D radiative transfer of the stellar radiation field in a gaseous nebula with the defined density distribution and chemical abundances. It produces the emission-line spectrum, the thermal structure and the ionization structure of the nebula. It allows us to determine the stellar characteristics and the nebula parameters. The atomic data sets used for the calculation are energy levels, collision strengths and transition probabilities from  the CHIANTI data base \citep[version 5.2;][]{Landi2006}, hydrogen and helium free--bound coefficients of \citet{Ercolano2006}, and opacities from \citet{Verner1993} and
\citet{Verner1995}. 

The best-fitting model was obtained through an iterative process, involving the comparison of the predicted H$\beta$ luminosity $L_{{\rm H}\beta}$(erg\,s${}^{-1}$), 
the flux intensities of some important lines, relative to H$\beta$ (such as $[$O~{\sc iii}$]$ $\lambda$5007 and  $[$N~{\sc ii}$]$
$\lambda$6584), with those measured from the observations.  The free parameters included distance and nebular parameters. We initially used the stellar luminosity ($L_{\star}=6000$\,L$_{\bigodot}$) and effective temperature ($T_{\rm eff}=70$kK) found by \citet{Todt2013}. However, we slightly adjusted the stellar luminosity to match the observed line flux of $[$O~{\sc iii}$]$ emission line. Moreover, we adopted the nebular density and abundances derived from empirical analysis in Section~\ref{a48:sec:empirical}, but they have been gradually adjusted until the observed nebular emission-line spectrum was reproduced by the model. The best-fitting $L_{{\rm H}\beta}$ depends upon the distance and nebula density. The plasma diagnostics yields $N_{\rm e} = 750$--1000\,cm$^{-3}$, which 
can be an indicator of the density range. Based on the kinematic analysis, the distance must be less than 2~kpc, but more than 1.5~kpc due to the large interstellar extinction. We matched the predicted H$\beta$ luminosity $L({\rm H}\beta)$ with the value derived from the observation by adjusting the distance and nebular density. Then, we adjusted abundances to get the best emission-line spectrum. 

\subsection{The ionizing spectrum}

\begin{figure}
\begin{center}
\includegraphics[width=3.3in]{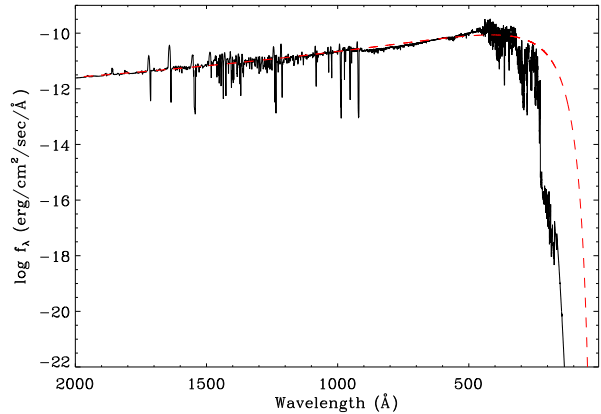}%
\caption{Non-LTE model atmosphere flux (solid line) calculated with the PoWR models for the surface abundances H:He:C:N:O~=~10:85:0.3:5:0.6 by mass and the stellar temperature $T_{\rm eff}$\,=\,70\,kK, compared with a blackbody (dashed line) at the same temperature.
}%
\label{a48:SED:PoWR}%
\end{center}
\end{figure}

The hydrogen-deficient synthetic spectra of Abell\,48 was modelled using stellar model atmospheres produced by
the Potsdam Wolf--Rayet (PoWR) models for expanding atmospheres \citep{Grafener2002,Hamann2004}. It solves the non-local thermodynamic equilibrium (non-LTE) radiative transfer equation in the comoving
frame, iteratively with the equations of statistical equilibrium and radiative equilibrium, for an expanding atmosphere
under the assumptions of spherical symmetry, stationarity and homogeneity. 
The result of our  model atmosphere is shown in Fig.\,\ref{a48:SED:PoWR}. The model atmosphere calculated with the PoWR code is for the stellar surface abundances H:He:C:N:O~=~10:85:0.3:5:0.6 by mass, the stellar temperature $T_{\rm eff}$\,=\,70\,kK, the transformed radius $R_{\rm t}=0.54$\,R${}_{\bigodot}$ and the wind terminal velocity $v_{\infty}=1000$~km\,s$^{-1}$. The best photoionization model was obtained with an effective temperature of 70\,kK \citep[the same as PoWR model used by][]{Todt2013} and a stellar luminosity of $L_{\rm \star}/$L$_{\bigodot}$=~5500, which is close to $L_{\star}/$L$_{\bigodot}$=~6000 adopted by \citet{Todt2013}. This stellar luminosity was found to be consistent with the observed H$\beta$ luminosity and the flux ratio of $[$O~{\sc iii}$]$/H$\beta$. 
A stellar luminosity higher than 5500\,L$_{\bigodot}$ produces inconsistent results for the nebular photoionization modelling. The emission-line spectrum produced by our adopted stellar parameters was found to be consistent with the observations.

\begin{table}
\caption{Input parameters for the \mbox{{\sc mocassin}} photoionization model. }
\label{a48:tab:mocassininput}
\centering
\begin{tabular}{lc|lcc}
\hline
\hline
\noalign{\smallskip}
\multicolumn{2}{c}{Stellar and Nebular} & & \multicolumn{2}{c}{Nebular Abundances} \\
\multicolumn{2}{c}{ Parameters}         &       &  Model        & Obs. 	        \\
\hline
$T_{\rm eff}$\,(kK)         	& 70    & He/H  & 0.120    		&  0.124 	     \\
$L_{\rm \star}$\,(L$_{\bigodot}$)& 5500 & C/H~$\times 10^{3}$   & 3.00	&  -- 	 \\
$N_{\rm H}$ (cm$^{-3}$)                	& 800-1200   & N/H~$\times 10^{5}$   & 6.50 	& 4.30   \\
$D$\,(kpc)                   	& 1.9   & O/H~$\times 10^{4}$   & 1.40 	& 1.59 	 \\
$r_{\rm out}$\,(arcsec)     	& 23  	& Ne/H~$\times 10^{5}$  & 6.00 	& 6.36	 \\
$\delta r$\,(arcsec)    		& 13 	& S/H~$\times 10^{6}$   & 6.00	& 6.73	 \\
$h$\,(arcsec)    				& 23    & Ar/H~$\times 10^{6}$  & 1.20 	& 1.48 	 \\
\hline
\end{tabular}
\end{table}

\subsection{The density distribution}

\begin{figure}
\begin{center}
\includegraphics[width=2.2in]{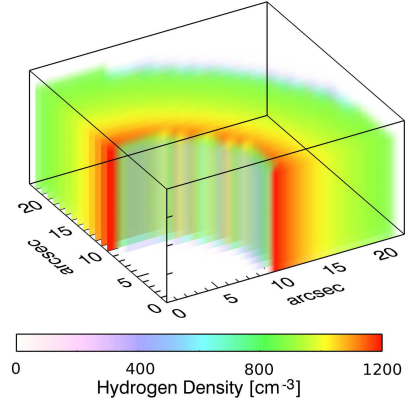}%
\caption{The density distribution based on the ISW models adopted for photoionization modelling of Abell~48. 
The cylinder has outer radius of $23\arcsec$ and thickness of $13\arcsec$. Axis units are arcsec, where 1 arcsec is equal to $9.30\times 10^{-3}$ pc based on the distance determined by our photoionization model.
\label{a48:fig10}%
}%
\end{center}
\end{figure}

We initially used a three-dimensional uniform density distribution, which was developed from our kinematic analysis. However, the interacting stellar winds (ISW) model developed by \citet{Kwok1978} demonstrated that a slow dense superwind from the AGB phase is swept up by a fast tenuous wind during the PN phase, creating a compressed dense shell, which is similar to what we see in Fig.~\ref{a48:fig10}. Additionally, \citet{Kahn1985} extended the ISW model to describe a highly elliptical mass distribution. This extension later became known as the generalized interacting stellar winds theory. There are a number of hydrodynamic simulations, which showed the applications of the ISW theory for bipolar PNe \citep[see e.g.][]{Mellema1996,Mellema1997}. As shown in Fig.~\ref{a48:fig10}, we adopted a density structure with a toroidal wind mass-loss geometry, similar to the ISW model. In our model, we defined a density distribution in the cylindrical coordinate system, which has the form $N_{\rm H}(r) = N_{0}[ 1 + (r/r_{\rm in})^{-\alpha} ],$
where $r$ is the radial distance from the centre, $\alpha$ the radial density dependence, $N_{0}$ the characteristic density, $r_{\rm in} = r_{\rm out}-\delta r$ the inner radius, $r_{\rm out}$ the outer radius and $\delta r$ the thickness.

The density distribution is usually a complicated input parameter to constrain. However, the values found from our plasma diagnostics ($N_{\rm e}=750$--1000 cm$^{-3}$) allowed us to constrain our density model. The outer radius and the height of the cylinder are equal to $r_{\rm out}=23\arcsec$ and the thickness is $\delta r=13\arcsec$. The density model and distance (size) were adjusted in order to reproduce $I$(H$\beta)=1.355 \times 10^{-10}$\,erg\,s$^{-1}$\,cm$^{-2}$, dereddened using c(H$\beta$)\,=\,3.1 (see Section~\ref{a48:sec:observations}). We tested distances, with values ranging from 1.5 to 2.0\,kpc. We finally adopted the characteristic density of $N_{0}=600$\,cm$^{-3}$ and the radial density dependence of $\alpha=1$. The value of 1.90\,kpc found here, was chosen, because of the best predicted  H$\beta$ luminosity, and it is in excellent agreement with the distance constrained by the synthetic spectral energy distribution (SED) from the PoWR models. Once the density distribution and distance were identified, the variation of the nebular ionic abundances were explored.

\begin{table}
\caption{Dereddened observed and predicted emission-line fluxes for Abell~48. References: D13 -- this work; T13 -- \citet{Todt2013}. Uncertain and very uncertain values are followed by `:' and `::', respectively. 
The symbol `*' denotes blended emission lines. 
\label{a48:tab:compute:lines}
}
\centering
\begin{tabular}{lccc}
\hline
\hline
\noalign{\smallskip}
Line			& \multicolumn{2}{c}{Observed} & {Predicted} \\
\noalign{\smallskip}
\cline{2-3}
\noalign{\smallskip}
						& D13		& T13		& 	\\
\noalign{\smallskip}
\hline
\noalign{\smallskip}
$I$(H$\beta$)/10$^{-10}\,\frac{\rm erg}{\rm cm^{2}s}$ 						
                        & 1.355  	& --		& 1.371		\\
\noalign{\smallskip}
H$\beta$ 4861			& 100.00 	& 100.00    & 100.00   	\\
H$\alpha$ 6563			& 286.00 	& 290.60    & 285.32   	\\
H$\gamma$ 4340			& 54.28: 	& 45.10  	& 46.88  	\\
H$\delta$ 4102			& -- 	    & --    	& 25.94   	\\
\noalign{\smallskip}
He~{\sc i} 4472			& 7.42: 	& --   		& 6.34   	\\
He~{\sc i} 5876			& 18.97 	& 20.60    	& 17.48   	\\
He~{\sc i} 6678			& 5.07  	& 4.80    	& 4.91  	\\
He~{\sc i} 7281			& 0.58:: 	& 0.70    	& 0.97   	\\
He~{\sc ii} 4686		& -- 		& --    	& 0.00   	\\
\noalign{\smallskip}
C~{\sc ii} 6462 		& 0.38 		& --    	& 0.27		\\
C~{\sc ii} 7236 		& 1.63 		& --    	& 1.90   	\\
\noalign{\smallskip}
$[$N~{\sc ii}$]$ 5755 	& 0.43:: 	& 0.40    	& 1.20   	\\
$[$N~{\sc ii}$]$ 6548 	& 26.09 	& 28.20    	& 26.60		\\
$[$N~{\sc ii}$]$ 6584	& 87.28 	& 77.00    	& 81.25		\\
\noalign{\smallskip}
$[$O~{\sc ii}$]$ 3726	& 128.96: 	& --    	& 59.96		\\
$[$O~{\sc ii}$]$ 3729	& *			& --    	& 43.54   	\\
$[$O~{\sc ii}$]$ 7320	& -- 		& 0.70    	& 2.16		\\
$[$O~{\sc ii}$]$ 7330	& -- 		& 0.60    	& 1.76		\\
$[$O~{\sc iii}$]$ 4363	& --		& 3.40    	& 2.30		\\
$[$O~{\sc iii}$]$ 4959	& 99.28 	& 100.50    & 111.82	\\
$[$O~{\sc iii}$]$ 5007	& 319.35 	& 316.50    & 333.66	\\
\noalign{\smallskip}
$[$Ne~{\sc iii}$]$ 3869	& 38.96  	& --    	& 39.60  	\\
$[$Ne~{\sc iii}$]$ 3967	& -- 		& --    	& 11.93		\\
\noalign{\smallskip}
$[$S~{\sc ii}$]$ 4069	& -- 		& --    	& 1.52   	\\
$[$S~{\sc ii}$]$ 4076	& -- 		& --    	& 0.52  	\\
$[$S~{\sc ii}$]$ 6717	& 7.44 		& 5.70    	& 10.30   	\\
$[$S~{\sc ii}$]$ 6731	& 7.99  	& 6.80    	& 10.57   	\\
$[$S~{\sc iii}$]$ 6312	& 0.60:: 	& --    	& 2.22   	\\
$[$S~{\sc iii}$]$ 9069	& 19.08 	& --    	& 16.37   	\\
\noalign{\smallskip}
$[$Ar~{\sc iii}$]$ 7136	& 10.88 	& 10.20    	& 12.75  	\\
$[$Ar~{\sc iii}$]$ 7751	& 4.00:: 	& --    	& 3.05   	\\
$[$Ar~{\sc iv}$]$ 4712	& -- 		& --    	& 0.61   	\\
$[$Ar~{\sc iv}$]$ 4741	& --		& --    	& 0.51   	\\
\noalign{\smallskip}
\hline
\end{tabular}
\end{table}

\subsection{The nebular elemental abundances}

\renewcommand{\baselinestretch}{0.9}
\renewcommand{\baselinestretch}{1.5}

\renewcommand{\baselinestretch}{0.9}
\renewcommand{\baselinestretch}{1.5}

Table~\ref{a48:tab:mocassininput} lists the nebular elemental abundances (with respect to H) used for the photoionization model. 
We used a homogeneous abundance distribution, since we do not have any direct observational evidence for the presence of chemical inhomogeneities.  
Initially, we used the abundances from empirical analysis as initial values for our modelling (see Section~\ref{a48:sec:empirical}). They were successively modified to fit the optical emission-line spectrum through an iterative process. We obtain a C/O ratio of 21 for Abell\,48, indicating that it is predominantly C-rich. Furthermore, we find a helium abundance of 0.12. This can be an indicator of a large amount of mixing processing in the He-rich layers during the He-shell flash leading to an increase carbon abundance. The nebulae around H-deficient CSs typically have larger carbon abundances than those with H-rich CSs \citep[see review by][]{DeMarco2001}. 
The ${\rm O}/{\rm H}$ we derive for Abell\,48 is lower than the solar value \citep[${\rm O}/{\rm H}=4.57\times 10^{-4}$;][]{Asplund2009}. This may be due to that the progenitor has a sub-solar metallicity. The enrichment of carbon can be produced in a very intense mixing process in the He-shell flash \citep{Herwig1997}. Other elements seem to be also decreased compared to the solar values, such as sulphur and argon. Sulphur could be depleted on to dust grains \citep{Sofia1994}, but argon cannot have any strong depletion by dust formation \citep{Sofia1998}. We notice that the N/H ratio is about the solar value given by \citet{Asplund2009}, but it can be produced by secondary conversion of initial carbon if we assume a sub-solar metallicity progenitor. The combined (C+N+O)/H ratio is by a factor of 3.9 larger than the solar value, which can be produced by multiple dredge-up episodes occurring in the AGB phase.

\renewcommand{\baselinestretch}{0.9}
\begin{figure*}
\begin{center}
\includegraphics[width=2.9in]{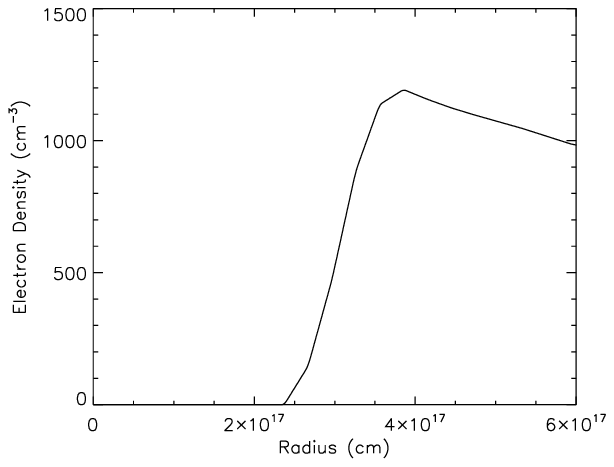}%
\includegraphics[width=2.9in]{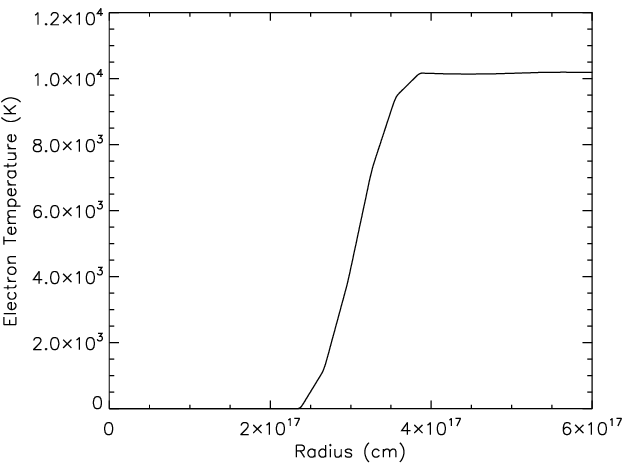}\\
\includegraphics[width=2.9in]{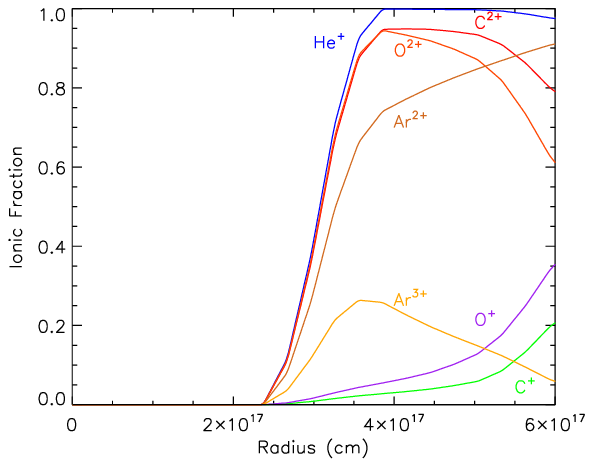}%
\includegraphics[width=2.9in]{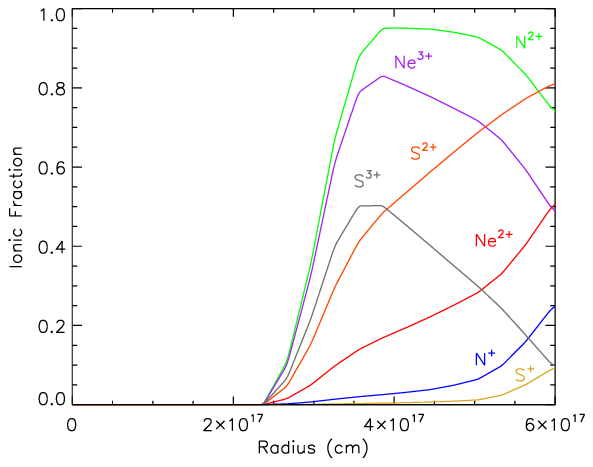}
\caption{Top: electron density and temperature as a function of radius along the equatorial direction. 
Bottom: ionic stratification of the nebula. Ionization fractions are shown for helium, carbon, oxygen, argon (left-hand panel), nitrogen, neon and sulphur (right-hand panel). 
\label{a48:ion:fraction}%
}%
\end{center}
\end{figure*}
\renewcommand{\baselinestretch}{1.5}

\renewcommand{\baselinestretch}{1.2}
\begin{table*}
\caption{Fractional ionic abundances for Abell\,48 obtained from the photoionization model.}
\label{a48:tab:ionfraction}
\centering
\begin{tabular}{lccccccc}
\hline  
\hline 
   & \multicolumn{7}{c}{Ion}\\
\cline{2-8}
Element & {\sc i}   &{\sc ii}   &{\sc iii}&{\sc iv}&{\sc v} &{\sc vi}&{\sc vii}\\
\hline 
H  & 3.84($-2$) & 9.62($-1$) &   &   &   &   &   \\ 
He & 3.37($-2$) & 9.66($-1$) & 1.95($-6$) &   &   &   &   \\ 
C  & 5.43($-4$) & 1.73($-1$) & 8.18($-1$) & 8.93($-3$) & 1.64($-15$) & 1.00($-20$) & 1.00($-20$) \\ 
N  & 1.75($-2$) & 1.94($-1$) & 7.79($-1$) & 8.98($-3$) & 2.72($-15$) & 1.00($-20$) & 1.00($-20$) \\ 
O  & 4.32($-2$) & 2.60($-1$) & 6.97($-1$) & 1.18($-7$) & 3.09($-20$) & 1.00($-20$) & 1.00($-20$) \\ 
Ne & 9.94($-3$) & 3.88($-1$) & 6.03($-1$) & 1.12($-13$) & 1.00($-20$) & 1.00($-20$) & 1.00($-20$) \\ 
S  & 6.56($-5$) & 8.67($-2$) & 6.99($-1$) & 2.12($-1$) & 2.42($-3$) & 1.66($-15$) & 1.00($-20$) \\ 
Ar & 2.81($-3$) & 3.74($-2$) & 8.43($-1$) & 1.17($-1$) & 1.02($-13$) & 1.00($-20$) & 1.00($-20$) \\ 
\hline 
\end{tabular}
\end{table*}
\renewcommand{\baselinestretch}{1.5}

\section{Model results}
\label{a48:sec:modelresults}

\subsection{Comparison of the emission-line fluxes}

Table~\ref{a48:tab:compute:lines} compares the flux intensities predicted by the best-fitting model with those from the observations. Columns 2 and 3 present the dereddened fluxes of our observations and those from \citet{Todt2013}. The predicted emission-line fluxes are given in Column 4, relative to the intrinsic dereddened H$\beta$ flux, on a
scale where $I($H$\beta)$\,=\,100. The most emission-line fluxes presented are in reasonable agreement with the observations. 
However, we notice that the [O~{\sc ii}]\,$\lambda$7319 and $\lambda$7330 doublets are overestimated by a factor of 3, which can be due to the recombination contribution. Our photoionization code incorporates the recombination term to the statistical equilibrium equations. However, the recombination contribution are less than 30 per cent for the values of $T_{\rm e}$ and $N_{\rm e}$ found from the plasma diagnostics. Therefore, the discrepancy between our model and observed intensities of these lines can be due to inhomogeneous condensations such as clumps and/or colder small-scale structures embedded in the global structure. It can also be due to the measurement errors of these weak lines. The [O~{\sc ii}]\,$\lambda\lambda$3726,3729 doublet predicted by the model is around 25 per cent lower, which can be explained by either the recombination contribution or the flux calibration error.
There is a notable discrepancy in the predicted [N~{\sc ii}]\,$\lambda$5755 auroral line, being higher by a factor of $\sim 3$. 
It can be due to the errors in the flux measurement of the [N~{\sc ii}]\,$\lambda$5755 line. The predicted [Ar~{\sc iii}]\,$\lambda$7751 line is also 30 per cent lower, while [Ar~{\sc iii}]\,$\lambda$7136 is about 20 per cent higher. The [Ar~{\sc iii}]\,$\lambda$7751 line usually is blended with the telluric line, so the observed intensity of these line can be overestimated. 
It is the same for [S~{\sc iii}]\,$\lambda$9069, which is typically affected by the atmospheric absorption band. 

\begin{table}
\caption{Integrated ionic abundance ratios for He, C, N, O, Ne, S and Ar, derived from model ionic fractions and compared to those from the empirical analysis.}
\label{a48:ion_compare}
\centering
\begin{tabular}{@{}lcc}
\hline
Ionic ratio			& Observed			& Model\\
\hline
He$^+$/H$^+$		& 0.124             & 0.116	\\
C$^{2+}$/H$^+$		& 2.16($-3$)		& 2.45($-3$)\\
N$^+$/H$^+$			& 1.42($-5$)		& 1.26($-5$)\\
O$^+$/H$^+$			& 5.25($-5$)		& 3.63($-5$)\\
O$^{2+}$/H$^+$		& 1.06($-4$)		& 9.76($-5$)\\
Ne$^{2+}$/H$^+$		& 4.26($-5$)		& 3.62($-5$)\\
S$^+$/H$^+$			& 3.98($-7$)		& 5.20($-7$)\\
S$^{2+}$/H$^+$		& 5.58($-6$)		& 4.19($-6$)\\
Ar$^{2+}$/H$^+$		& 9.87($-7$) 		& 1.01($-6$)\\
\hline
\end{tabular}
\end{table}

\subsection{Ionization and thermal structure}

The volume-averaged fractional ionic abundances are listed in Table~\ref{a48:tab:ionfraction}. We note that hydrogen and helium are singly-ionized. We see that the O$^{+}$/O ratio is higher than the N$^{+}$/N ratio by a factor of 1.34, which is dissimilar to what is generally assumed in the $icf$ method. However, the O$^{2+}$/O ratio
is nearly a factor of 1.16 larger than the Ne$^{2+}$/Ne ratio, in agreement with the general assumption for $icf$(Ne). We see that only 19 per cent  of the total nitrogen in the nebula is in the form of N$^{+}$. However, the total oxygen largely exists as O$^{2+}$ with 70 per cent and then O$^{+}$ with 26 per cent.

The elemental abundances we used for the photoionization model returns ionic abundances listed in Table~\ref{a48:ion_compare}, are comparable to those from the empirical analysis derived in Section~\ref{a48:sec:empirical}. The ionic abundances derived from the observations do not show major discrepancies in He$^{+}$/H$^{+}$, C$^{2+}$/H$^{+}$, N$^{+}$/H$^{+}$, O$^{2+}$/H$^{+}$, Ne$^{2+}$/H$^{+}$ and  Ar$^{2+}$/H$^+$; differences remain below 18 per cent.  However, the predicted and empirical values of O$^{+}$/H$^{+}$, S$^{+}$/H$^+$ and S$^{2+}$/H$^+$ have discrepancies of about 45, 31 and 33 per cent, respectively.

Fig.~\ref{a48:ion:fraction}(bottom) shows plots of the ionization structure of helium, carbon, oxygen, argon (left-hand panel), nitrogen, neon and sulphur (right-hand panel) as a function of radius along the equatorial direction. As seen, ionization layers have a clear ionization sequence from the highly ionized inner parts to the outer regions. Helium is 97 percent singly-ionized over the shell, while oxygen is 26 percent singly ionized and 70 percent doubly ionized. Carbon and nitrogen are about $\sim20$ percent singly ionized $\sim80$ percent doubly ionized. The distribution of N$^{+}$ is in full agreement with the IFU map, given in Fig\,\ref{a48:ifu:ion:map}. Comparison between the He$^{+}$, O$^{2+}$ and S$^{+}$ ionic abundance maps obtained from our IFU observations and the ionic fractions predicted by our photoionization model also show excellent agreement.

\begin{table}
\caption{Mean electron temperatures (K) weighted by ionic species for the whole nebula obtained from the photoionization model.}
\label{a48:tab:temperatures}
\centering
\begin{tabular}{lccccccc}
\hline
\hline
& \multicolumn{7}{c}{Ion}\\
\cline{2-8}
  El. & {\sc i}   &{\sc ii}   &{\sc iii}&{\sc iv}&{\sc v} &{\sc vi}&{\sc vii}\\
\hline
H &     9044 &    10194 &   &   &   &   &   \\ 
He &     9027 &    10189 &    10248 &   &   &   &   \\ 
C &     9593 &     9741 &    10236 &    10212 &    10209 &    10150 &    10150 \\ 
N &     8598 &     9911 &    10243 &    10212 &    10209 &    10150 &    10150 \\ 
O &     9002 &    10107 &    10237 &    10241 &    10211 &    10150 &    10150 \\ 
Ne &     8672 &    10065 &    10229 &    10225 &    10150 &    10150 &    10150 \\ 
S &     9386 &     9388 &    10226 &    10208 &    10207 &    10205 &    10150 \\ 
Ar &     8294 &     9101 &    10193 &    10216 &    10205 &    10150 &    10150 \\ 
\hline
\end{tabular}
\end{table}

Table~\ref{a48:tab:temperatures} lists mean temperatures weighted by the ionic abundances. 
Both [N~{\sc ii}] and [O~{\sc iii}] doublets, as well as He\,{\sc i} lines arise from the same ionization zones, so they should have roughly similar values. The ionic temperatures increasing towards higher ionization stages could also have some implications for the mean temperatures averaged over the entire nebula. However, there is a large discrepancy by a factor of 2 between our model and ORL empirical value of $T_{\rm e}$(He\,{\sc i}$)$. This could be due to some temperature fluctuations in the nebula  \citep{Peimbert1967,Peimbert1971}. The temperature fluctuations lead to overestimating the electron temperature deduced from CELs. This can lead to the discrepancies in abundances determined from CELs and ORLs \citep[see e.g.][]{Liu2000}. 
Nevertheless, the temperature discrepancy can also be produced by bi-abundance models \citep{Liu2003,Liu2004b}, containing some cold hydrogen-deficient material, highly enriched in helium and heavy elements, embedded in the diffuse warm nebular gas of normal abundances. The existence and origin of such inclusions are still unknown. It is unclear whether there is any link between the assumed H-poor inclusions in PNe and the H-deficient CSs. 

\section{Conclusion}
\label{a48:sec:conclusions}

We have constructed a photoionization model for the nebula of Abell~48. This consists of a dense hollow cylinder, assuming homogeneous abundances. The three-dimensional density distribution was interpreted using the morpho-kinematic model determined from spatially resolved kinematic maps and the ISW model.  Our aim was to construct a model that can reproduce the nebular emission-line spectra, temperatures and ionization structure determined from the observations. We have used the non-LTE model atmosphere from \citet{Todt2013} as the ionizing source. Using the empirical analysis methods, we have determined the temperatures and the elemental abundances from CELs and ORLs. We notice a discrepancy between temperatures estimated from $[$O~{\sc iii}$]$ CELs and those from the observed He\,{\sc i} ORLs. In particular, the abundance ratios derived from empirical analysis could also be susceptible to inaccurate values of electron temperature and density. However, we see that the predicted ionic abundances are in decent agreement with those deduced from the empirical analysis. The emission-line fluxes obtained from the model were in fair agreement with the observations.  

We notice large discrepancies between He\,{\sc i} electron temperatures derived from the model and the empirical analysis. The existence of clumps and low-ionization structures could solve the problems \citep{Liu2000}. Temperature fluctuations have been also proposed to be responsible for the discrepancies in temperatures determined from CELs and ORLs \citep{Peimbert1967,Peimbert1971}. Previously, we also saw large ORL--CEL abundance discrepancies in other PNe with hydrogen-deficient CSs, for example Abell~30 \citep{Ercolano2003b} and NGC~1501 \citep{Ercolano2004}. 
A fraction of H-deficient inclusions might produce those discrepancies, which could be ejected from the stellar surface during a very late thermal pulse (VLTP) phase or born-again event \citep[]{Iben1983}. However, the VLTP event is expected to produce a carbon-rich stellar surface abundance \citep{Herwig2001}, whereas in the case of Abell~48 negligible carbon was found at the stellar surface \citep[C/He~=~$3.5\times10^{-3}$ by mass;][]{Todt2013}. The stellar evolution of Abell~48 still remains unclear and needs to be investigated further. But, its extreme helium-rich atmosphere (85 per cent by mass) is more likely connected to a merging process of two white dwarfs as evidenced for
R Cor Bor stars of similar chemical surface composition by observations \citep{Clayton2007,Garcia-Hernandez2009} and hydrodynamic simulations \citep{Staff2012,Zhang2012,Menon2013}. 

We derived a nebula ionized mass of $\sim0.8$~M$_{\bigodot}$. The high C/O ratio indicates that it is a predominantly C-rich nebula. The C/H ratio is largely over-abundant compared to the solar value of \citet{Asplund2009}, while oxygen, sulphur and argon are under-abundant. Moreover, nitrogen and neon are roughly similar to the solar values. Assuming a sub-solar metallicity progenitor, a 3rd dredge-up must have enriched carbon and nitrogen in AGB-phase. However, extremely high carbon must be produced through mixing processing in the He-rich layers during the He-shell flash. The low N/O ratio implies that the progenitor star never went through the hot bottom burning phase, which occurs in AGB stars with initial masses more than 5M$_{\bigodot}$ \citep{Karakas2007,Karakas2009}. Comparing the stellar parameters found by the model, $T_{\rm eff}$\,=\,70\,kK and $L_{\rm \star}/$L$_{\bigodot}$=~5500, with VLTP evolutionary tracks from \citet{Bloecker1995b}, we get a current mass of $\sim 0.62 {\rm M}_{\bigodot}$, which originated from a progenitor star with an initial mass of $\sim 3 {\rm M}_{\bigodot}$. However, the VLTP evolutionary tracks by \citet{MillerBertolami2006a} yield a current mass of $\sim 0.52 {\rm M}_{\bigodot}$ and a progenitor mass of $\sim1{\rm M}_{\bigodot}$, which is not consistent with the derived nebula ionized mass. Furthermore, time-scales for VLTP evolutionary track \citep{Bloecker1995b} imply that the CS has a post-AGB age of about $\sim$\,9\,000 yr, in agreement with the nebula's age determined from the kinematic analysis. We therefore conclude that Abell~48 originated from an $\sim3$~M$_{\bigodot}$ progenitor, which is consistent with the nebula's features.

\section*{Acknowledgments}

AD warmly acknowledges the award of an international Macquarie University
Research Excellence Scholarship (iMQRES). BE is supported by the German Research Foundation (DFG) Cluster of Excellence ``Origin and Structure of the Universe''. AYK acknowledges the support from the National Research Foundation (NRF) of South Africa. We would like to thank Prof. Wolf-Rainer Hamann, Prof. Simon Jeffery and Dr. Amanda Karakas for illuminating discussions and helpful comments.  We would also like to thank Dr. Kyle DePew for carrying out the 2010 ANU 2.3 m observing run. AD thanks Dr. Milorad Stupar for assisting with the 2012 ANU 2.3 m observing run and his guidance on the \textsc{iraf} pipeline {\sc wifes}, Prof. Quentin A. Parker and Dr. David J. Frew for helping in the observing proposal writing stage, and the staff at the ANU Siding Spring Observatory for their support. We would also like to thank the anonymous referee for helpful suggestions.

\label{lastpage}

\end{document}